\documentclass[journal]{IEEEtran}
\usepackage{filecontents}
\usepackage{amsmath,amsthm,dsfont,tabularx,scalefnt,comment}
\usepackage[cmintegrals]{newtxmath}
\usepackage[nocompress]{cite}
\usepackage{subcaption}
\usepackage{caption}
\usepackage{bm}
\usepackage{color,soul}
\usepackage{multirow}
\usepackage{siunitx}
\usepackage{setspace,adjustbox}

\usepackage{algorithm}
\usepackage{algpseudocode}

\usepackage{amsmath,amsthm}
\usepackage{mathtools}

\begin{document}

\bibliographystyle{unsrt}

\title{Quantum-enhanced Information Retrieval from Reflective Intelligent Surfaces}
\author{
\IEEEauthorblockN {
   Shiqian Guo, 
   Tingxiang Ji,
   Jianqing Liu
}

\IEEEauthorblockA {
  Department of Computer Science, North Carolina State University, Raleigh, USA 27606.\\
}
\IEEEauthorblockA {
    \{sguo26, tji2, jliu96\}@ncsu.edu
}

\thanks{This work is supported in part by the National Science Foundation under grants 2304118 and 2326746.}
}

\maketitle


\begin{abstract}
Information retrieval from passive backscatter systems is widely used in digital applications with tight energy budgets, short communication distances, and low data rates. Due to the fundamental limits of classical wireless receivers, the achievable data rate cannot be increased without compromising either energy efficiency or communication range, thereby hindering the broader adoption of this technology. In this work, we present a novel time-resolving quantum receiver combined with a multi-mode probing signal to extract large-alphabet information modulated by a passive reconfigurable intelligent surface (RIS). The adaptive nature of the proposed receiver yields significant quantum advantages over classical receivers without relying on complex or fragile quantum resources such as entanglement. Simulation results show that the proposed technique surpasses the classical standard quantum limit (SQL) for modulation sizes up to $M = 2^8$, meanwhile halving the probing energy or increasing the communication distance by a factor of 1.41.
\end{abstract}

\IEEEpeerreviewmaketitle

\section{Introduction}
Backscatter communication, which relies on passive reflection and modulation of incident wireless signals, has played a critical role in low-power, low-complexity and ambient data storage and retrieval. Nowadays, many backscatter devices such as barcodes and RFID tags are found in various applications including inventory management, precision agriculture, smart buildings, and large-scale industrial automation \cite{aboagye2022ris}. A backscatter system typically consists of a passive backscatter device and a reader. The backscatter device can be considered as a digital memory that encodes data in different degrees of freedom (DoF) such as amplitude (as in CDs and DVDs), space (as in barcodes), and frequency and phase (as in chipless RFIDs). During information retrieval, the reader transmits a signal to the backscatter device and receives a reflected response. When the reader and backscatter device are separated by a distance of d, the returned signal is often weak because of the $d^{-4}$ path loss factor (multiplicative loss of the forward and backward channels). Together, the limited DoF imposed by the physical aperture of the backscatter device and the inherently low signal-to-noise ratio (SNR) at the reader pose a fundamental constraint on the number of bits that can be reliably encoded and retrieved. Consequently, contemporary backscatter systems remain restricted to short-range, low-data-rate applications.

To break the fundamental rate-range limit of backscatter systems, a reader with quantum sensing capability holds the promise. Compared with classical sensing technologies whose sensitivity is constrained by the shot-noise limit, also known as the standard quantum limit (SQL) due to Gaussian noises in independent measurements, quantum sensing is a revolutionary technology for detecting ultraweak signals with sensitivity unmatched by classical sensing \cite{liu2024road}. It exploits non-classical resources such as entanglement, superposition, and unconventional receivers to surpass SQL and ultimately approach the Heisenberg limit (HL) that is only dictated by physics. In existing studies, researchers have demonstrated in both theory and experiments the advantage of quantum sensing to retrieve information from barcodes, QR codes, and other types of backscatter devices\cite{harney2021idler,banchi2020quantum}. Most of these works realize the quantum advantage by using entangled photons or squeezed light as the probing signal to impinge the backscatter device. While the quantum probing light significantly enhances the sensing accuracy, it is a fragile resource that is hard to generate and prone to decoherence in long-range transmissions. In light of it, this paper presents the design of a new quantum sensing-enabled backscatter system. It is based on simple coherent probing light (e.g., classical laser) and an adaptive time-resolving quantum receiver. The backscatter device of concern is an RIS that encodes data in a much larger DoF than most backscatter devices. In this challenging reading scenario, the paper reports that the design still manages to beat SQL, saves transmit power, and extends the communication distance without using entanglement resources. 

To give a brief overview of our design, the proposed probing light consists of multiple modes (i.e., multiple colors or wavelengths) and impinges on the RIS. The RIS modulates the information pre-stored in an optical memory by adjusting the phase and amplitude of the impinged light; and subsequently focuses and directs the light to the receiver. The receiver consists of a single-photon detector preceded by an adaptive displacement module that extracts information from the incoming optical signal. The displacement module functions as a mixer, using a local oscillator (LO) signal to interfere with the received light. When the LO signal perfectly matches the incoming signal, the displacement operation ideally maps the input state to the vacuum, resulting in no photon detections at the output. In this ideal scenario, the detection of even a single photon rules out the corresponding hypothesis. After each detection event, the receiver updates the most likely hypothesis accordingly. We propose a Bayesian algorithm to search for optimized LO signal adaptively. It considers photon arrival times of different modes and selects the one that yields the maximum posterior probability to update the posterior distribution. The optimal LO signal for the next measurement step is then chosen to maximize this posterior. In summary, the main contributions of this work are summarized as follows:
\begin{itemize}
\item We investigate a RIS-based backscatter system, where the RIS performs both amplitude and phase modulation to enable large-alphabet encoding. A quantum sensing-based reader is developed for information retrieval whose accuracy surpasses the SQL without using entanglement resources. 
\item We introduce a novel adaptive time-resolving quantum receiver in conjunction with a multi-color probing light. This design extends the reading distance, simplifies the system complexity while efficiently extracting the maximum information from the RIS-modulated light.
\item Extensive simulations verify that the proposed design surpasses the classical SQL for modulation sizes up to $M = 2^8$. Moreover, the design requires only half the energy (or equivalently, enabling $\sqrt{2}$-times longer distances) of classical techniques to achieve the same target accuracy, implying strong practical quantum advantages. 
\end{itemize}

The remainder of this paper is organized as follows. Section \ref{sec:related} surveys recent works on quantum enhanced information retrieval. Section \ref{sec:system} presents the system model and describes the RIS modulation scheme together with the adaptive time-resolving quantum receiver. Section \ref{sec:results} discusses the simulation results, and Section \ref{sec:conclude} concludes the paper.

\section{Related Works}\label{sec:related}
\subsection{Backscatter-based Information Retrieval}
RIS is also known as programmable metasurface, which consists of repetitive and tunable metamaterial elements and is controlled by digital modules such as field-programmable gate arrays (FPGAs). It can dynamically and flexibly manipulate the properties of incoming wireless signals (e.g., microwave, THz, or visible light) including the amplitude, phase, polarization, and frequency. RIS becomes increasingly important in modern wireless systems, typically in three use cases. The first is the RIS-based simplified transceiver architecture that directly realizes signal modulations without using complicated digital-analog converters or mixers \cite{cheng2022reconfigurable}. The second is RIS-assisted wireless environment modulations, which means improving communication quality by beam manipulation (e.g., avoiding blockage) \cite{du2023semantic,zhou2024modulation}. The third is to encode information in RIS for high-capacity and programmable data encoding and high-efficiency retrieval (e.g., metasurface holography) compared to classical RFID systems \cite{ni2013metasurface,feng2024ris}. In this work, we focus on the use of RIS as a data encoder and investigate how to engineer a quantum transmitter and receiver for high-capacity information retrieval. 

When it comes to using RIS as a backscatter device, many existing works have studied it from different perspectives \cite{du2023semantic, zhou2024modulation, feng2024ris}. In \cite{du2023semantic}, Du employed RIS for analog modulation to enable signal compression. In \cite{zhou2024modulation, feng2024ris}, Zhou and Feng utilized RIS for digital symbol modulation. By leveraging RIS-assisted modulation, the receiver benefits from stronger and more structured signals, higher information throughput and spectral efficiency, and improved resource efficiency in wireless communications. Backscatter communications in the visible light wavelength is similar to the RF wavelength. RIS in visible light wavelength is proven capable to optimize both communication and illumination performance \cite{aboagye2022ris}. Specifically, for enhanced illumination, RIS can act as a controllable optical surface that adaptively modifies the system’s field of view (FoV) through coherent field addition, enabling either high-intensity focusing or wide-coverage illumination as needed. In this way, RIS provides a programmable FoV, supporting both beam-steering and beam-narrowing without physically moving optical components.

\subsection{Quantum Reading: Transmitter Design}
Quantum-enhanced information retrieval, also known as quantum reading, has lately attracted significant research attention and was first introduced by Pirandola et al. \cite{pirandola2011quantum}. In this framework, information retrieval from an optical memory is modeled as a problem of bosonic channel discrimination \cite{pirandola2021quantum}. The problem of quantum channel discrimination has a very rich theoretical structure due to its inherent double-optimization nature: one must jointly optimize both the input probe states and the output measurements. 

Zhuang and Pirandola \cite{zhuang2020ultimate} derived the ultimate limits for multiple quantum channel discrimination, providing fundamental lower bounds on the error probability achievable by any protocol. Their work is fully theoretical and considers optimization over both input states and measurement strategies. Similarly, Bergh et al. \cite{bergh2024parallelization} contributed to the theoretical foundations of quantum reading by showing that general adaptive channel discrimination strategies can be efficiently reduced to non-adaptive strategies with negligible performance loss, and that the corresponding discrimination performance admits a polynomial-time-computable upper bound even in the non-asymptotic regime. Rossi, Yu, and Chuang \cite{rossi2022quantum} investigated noisy quantum channel discrimination and studied strategies that optimize the input probe state, including entangled states, while assuming the receiver performs the optimal measurement given that input. Their work highlights how careful engineering of the input state alone can yield significant quantum advantage, even in the presence of noise.  Furthermore, Harney and Pirandola \cite{harney2021idler} showed that, in the context of quantum reading and bosonic multi-channel discrimination, optimal performance can be achieved without idler modes. By carefully designing multipartite probe states, the input state alone can be optimized to attain the same discrimination performance that would otherwise require idler-assisted adaptive strategies, highlighting the central role of input state engineering in quantum reading protocols. 

\subsection{Quantum Reading: Receiver Design}
Although optimizing the input state can significantly enhance quantum channel discrimination, this perspective remains largely theoretical. To achieve more practical improvements in discrimination performance, recent research has increasingly focused on optimizing the measurement strategies, aiming to design more effective receiver architectures. Static interferometric receivers with fixed displacements and photon-counting detectors have been demonstrated and analyzed for multi-symbol coherent-state discrimination, yielding lower symbol-error probabilities \cite{sidhu2021quantum,sidhu2023linear}. 
Complementary approaches aimed at reducing the fundamental noise penalty of heterodyne detection --- classical receiver design --- show that judicious use of correlated detection channels or nonclassical ancillae can suppress image-band or cross-spectral noise contributions, thereby improving SNR in phase-insensitive heterodyne readout without real-time control\cite{feng2022quantum,anai2024quantum}. Taken together, these results motivate quantum-reading designs in which the measurement policy itself is optimized for enhanced channel discrimination. 

While feedback-free architectures in the above literature offer practical simplicity, adaptive feedback protocols remain the benchmark for achieving the ultimate quantum limits HL in channel discrimination. By dynamically updating the receiver settings (e.g., displacement amplitudes, phase shifts, or local oscillator strengths) based on the outcomes of prior measurements, adaptive receivers can effectively approximate the HL, thereby minimizing the probability of error beyond the capability of any classical receivers \cite{le2025distributed,guo2025two}.
Becerra et al. \cite{becerra2013experimental,becerra2015photon} demonstrated an adaptive displacement receiver with real-time feed-forward that surpasses the SQL for multiple nonorthogonal coherent states. Moreover, by incorporating photon-number-resolving detectors, they showed that the adaptive strategy can be significantly enhanced and made robust against realistic imperfections such as limited visibility and excess noise. Ferdinand et al.\cite{ferdinand2017multi} generalized the adaptive displacement measurement strategy to multi-state coherent states and showed experimentally that it can beat the SQL at single-photon levels, something traditional binary LO adaptive displacement methods cannot do. Besides the photon-number-resovling receiver, Müller and Marquardt \cite{muller2015robust} developed a robust receiver for phase-shift-keyed (PSK) coherent states that exploits photon arrival-time information, enabling reliable discrimination under realistic operating conditions. Similarly, by tracking phase and intensity noise, the time-resolving quantum receiver proposed by Suo \cite{suo2025phase} can maintain its enhanced discrimination performance even in the presence of noise.

\subsection{Quantum Reading for Modulated Information}
Large-alphabet quantum reading has recently attracted significant attention because it enables high spectral efficiency and dense information encoding.
Burenkov \cite{burenkov2020time} demonstrates that combining a CFSK (coherent frequency-shift keying) modulation protocol with a time-resolving receiver is particularly advantageous for discriminating nonorthogonal coherent states at low average photon numbers, even for communication alphabets of four bits per symbol. Jabir \cite{jabir2022energy} further investigates the performance of matched modulation and receiver measurements for resource-efficient, quantum-enhanced optical communication. However, these experimental advances in quantum measurement have primarily addressed low-modulation-state discrimination problems, leaving open the challenge of extending such strategies to higher-dimensional or multi-bit encoding scenarios relevant for quantum reading and general channel discrimination.

\section{Quantum-enhanced Information Retrieval from RIS}\label{sec:system}
\subsection{System Overview}

\begin{figure}[htpb]
	\centering
	\subfloat{
		\includegraphics[width=\linewidth]{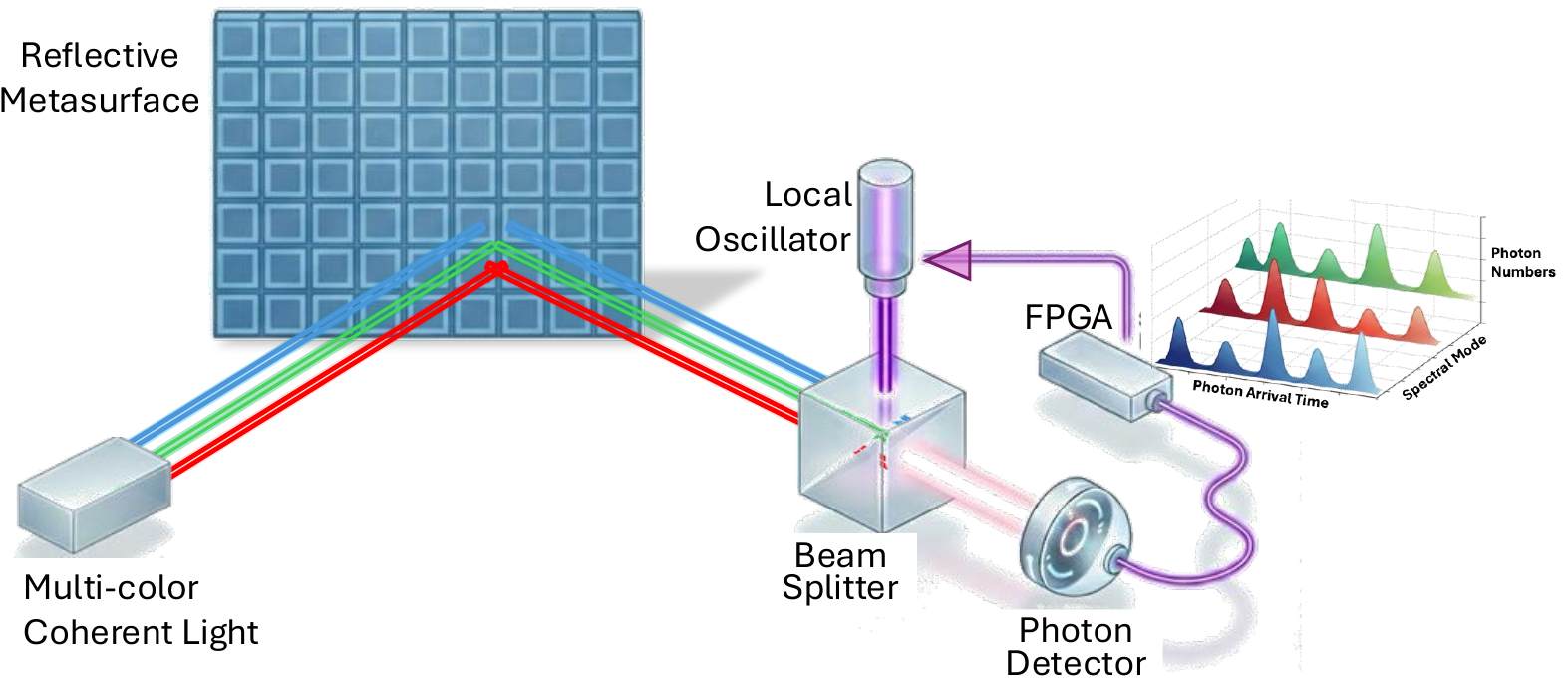}}
	\caption{RIS-modulated information retrieval system with an adaptive time-resolving quantum receiver.}
    \label{fig:RIS-QR-system}
\end{figure} 
As illustrated in Fig.~\ref{fig:RIS-QR-system}, we consider a free-space optical backscatter system equipped with a novel time-resolving quantum receiver. The system can operate in any optical channels, but to avoid ambient noises from visible light, we assume that the system operates in a less interfering spectrum such as infrared. There are three main components in the backscatter system: a classical laser source, an RIS, and a quantum receiver. Specifically, the quantum receiver consists of four modules --- a classical laser or LO, a beam splitter (BS), a single-photon detector (SPD), and a FPGA processing and control unit. Collectively, they form a closed feedback loop for adaptive measurement. 

To begin with, we present an overview of how the system works. The laser source first emits a bundle of  probing light of different spectral modes (or colors) toward the RIS. Each mode is a coherent light (or state) with the same phase, wavelength, and amplitude. The RIS is a passive reflective metasurface with a digital memory storing information of interest. When the multi-color probing light impinges the RIS, each mode is modulated to carry a set of states $\alpha_m$'s where the $m$-th symbol is preloaded in RIS's digital memory. Specifically, we let all probing modes be modulated with the same symbols to facilitate information retrieval at the receiver. In classical heterodyne detection --- widely seen in modern digital receivers, the LO is fixed and not adaptive. Therefore, its amplitude and phase do not influence the measurement statistics. Therefore, the measurement outcome of the laser source mode $s$ in a classical heterodyne receiver can be modeled as
\begin{equation}
\vartheta = \sqrt{\xi_{s} \eta_{s}}\, \alpha_m + \delta, 
\qquad \delta \sim \mathcal{CN}(0,1),
\end{equation}
where $\alpha_m$ is the transmitted coherent state amplitude, $\xi_{s} \eta_{s}$ is the total channel and detection efficiency for laser source mode $s$, and $\delta$ represents circularly symmetric complex Gaussian noise with unit variance, which includes the contribution of the vacuum LO.

Given a measurement outcome $\vartheta$, the receiver evaluates the distance to all possible symbols and selects the closest one:
\[
m^{*} = \arg\min_{m'} \left| \vartheta - \sqrt{\xi_{s} \eta_{s}}\,\alpha_{m'} \right|^2,
\]
which corresponds to the maximum likelihood decision rule under Gaussian noise. The detected symbol is thus $\alpha_{m^*}$, obtained through this classical state-discrimination measurement. The average symbol error probability of the system is estimated via Monte Carlo simulation as
\[
P_e = \frac{1}{N_m} \sum_{k=0}^{N_m-1} \mathbb{I}(\alpha_m, \alpha_{m^*}),
\]
where $N_m$ is the total number of transmitted symbols, and the indicator function $\mathbb{I}(\cdot,\cdot)$ is defined as
\[
\mathbb{I}(x_1, x_2) =
\begin{cases}
0, & x_1 = x_2,\\[3pt]
1, & x_1 \ne x_2.
\end{cases}
\]
Due to the Gaussian noise, the classical heterodyne receiver has a fundamental precision limited dictated by SQL. 
SQL represents the minimum probability of estimation error of an unknown parameter achievable using classical measurements without using any quantum resources such as entanglement. 

In contrast to classical measurement strategies, the proposed quantum receiver applies a displacement operation, realized by the BS that mixes the LO and incoming signals, using a run-time adaptive LO signal $\beta$. The incoming signal and the LO signal are assumed to arrive at the BS simultaneously with precise synchronization. The LO signal maintains a constant amplitude and phase throughout a single measurement shot and is updated for the next measurement shot. In this setup, the mean photon number of the arrival symbol after mixing with a fixed LO signal $\beta$ is $\{\langle n(m,\beta) \rangle\}_{s=1}^{S}$.

\begin{align}\label{eq:mean-photon}
\langle n(m,\beta) \rangle
=
\frac{\xi \eta}{T} \Big[
|\alpha_m|^2 + |\beta|^2 
- 2 V |\alpha_m| |\beta| \cos\big(\angle \alpha_m - \angle \beta \big)
\Big],
\end{align}
where $V$ is the interference visibility of the LO, and $T$ is the symbol duration (assuming no temporal spread after channels). We can deduce that if the LO signal and incoming signal are in tune in phase and amplitude, the mean photon number is near zero when $V \approx 1$, creating a vacuum state in the SPD.

By further assuming time-independent photon statistics, the number of photons detected at the SPD follows a Poisson distribution with mean $\langle n(m,\beta) \rangle$. Since each mode at the source is transmitted using a time-windowed pulse of duration $T$, the mean photon number $\langle n(m,\beta) \rangle$ of the displaced signal is linearly proportional to the relative time $\tau_s$ elapsed since the last measurement step, i.e., $n(\tau_s) \sim$ Poisson($\langle n(m,\beta) \rangle \tau_s$). Consequently, the probability density for detecting a photon exactly at a relative time $\tau_s$ is given by the temporal derivative of the non-vacuum probability $P(n(\tau_s) \neq 0)$ \cite{muller2015robust}:
\begin{equation}\label{eq:photon-arrival-time}
    \mathcal{L}_s(\tau_{s})
    = \frac{\partial}{\partial \tau_{s}}\!\left( 1 - e^{-\langle n(m,\beta) \rangle \tau_{s}} \right)
    = {\langle n(m,\beta) \rangle} e^{-\langle n(m,\beta) \rangle \tau_{s}}.
\end{equation}
Provided with the photon-arrival-time information at the SPD, we can incorporate it into the likelihood function which enables a Bayesian adaptive displacement strategy to adjust the LO signal, thus altering $\langle n(m,\beta) \rangle$ for the next measurement. In so doing, the posterior probability of the transmitted
hypothesis is updated in each round till the end of symbol duration. Intuitively, an ideal outcome is that the LO signal has the same amplitude and phase as the incoming light, effectively reproducing the correct symbol. 



\subsection{Multi-color Probing Light}
We assume that each coherent state is transmitted using a rectangular-window pulse 
with duration $T$. 
The transmitted amplitude may be scaled to account for the channel efficiency $\xi_s$ and the detection efficiency $\eta_s$. 
Let $S$ be the number of spectral modes, each of which is a coherent state with the same identical amplitude 
$\alpha_{0} = \sqrt{n_{0}/S}$ but at a distinct optical wavelength. $n_{0}$ is the overall photon intensity of the laser source, and it is evenly divided into $S$ modes.  
Because the sources are mutually incoherent, their instantaneous electric 
fields simply superpose. The total received field under the rectangular-window pulse 
envelope is therefore
\begin{equation*}
    E(t) = u(t)\sum_{s=1}^{S} \alpha_{0}\, e^{j 2\pi f_{s} t},
\end{equation*}
where $u(t) = 1$ for $0 \le t \le T$ and $u(t) = 0$ otherwise, $f_{s} = c / \lambda_{s}$ is the optical frequency corresponding to wavelength $\lambda_{s}$, and $c$ denotes the speed of light.

\subsection{RIS-Based Modulation}
The RIS is configured to encode the bits stored in its digital memory by jointly modulating the amplitude and phase of the incoming multi-color probing light.  
Assuming that the distance between source and RIS is much larger than RIS's aperture, the path-length differences across RIS elements are negligible. Under this far-field condition, only the programmable RIS phase profile contributes to the received complex field, and the input coherent-state energy is uniformly distributed over the illuminated $K$ RIS elements.

When the multi-color probing light impinges the RIS, the total received field corresponding to the $m$-th symbol is given by the coherent sum over the RIS elements and the temporal carrier of each wavelength component:
\begin{equation*}
E_{m}(t)
= 
u(t)\sum_{s=1}^{S}
\left(
 \sum_{k=1}^{K}
   \rho_k \, \frac{\alpha_0}{\sqrt{K}} \, e^{j\psi_m}
\right)
e^{j 2\pi f_s t},
\label{eq:E_simple}
\end{equation*}
where $\rho_k \in \{0,1\}$ denotes the binary amplitude-modulation coefficient of the $k$th RIS element, indicating whether the element is switched ``off'' ($0$) or ``on'' ($1$),  
$\psi_m$ is the modulated phase. Since the RIS performs coherent field summation across its $K$ elements, the effective coherent-state amplitude associated with the $m$-th encoded symbol becomes
\begin{equation*}
\alpha_m
=
\left|
\sum_{k=1}^{K}
\rho_k \frac{\alpha_0}{\sqrt{K}}
\right|
e^{j\psi_m},
\quad m = 1,2,\dots,M,
\end{equation*}
which reflects the RIS focusing gain arising from constructive field addition.  
Importantly, coherent lasers with distinct optical frequencies remain mutually incoherent even after RIS focusing, and therefore only same-wavelength components interfere coherently.

\subsection{Channel and Noise Model}
We model the source-RIS and RIS-receiver channels using the scalar diffraction theory. 

\textbf{Source-to-RIS channel:} The effective beam width at the RIS is determined by the source aperture:
\[
\theta_{\mathrm{source}} \sim \frac{\lambda_{s}}{\sqrt{A_{\mathrm{Tx}}}},
\]
where \(\lambda_{s}\) is the wavelength of the mode $s$ and \(A_{\mathrm{Tx}}\) is the transmitter aperture area.  
Assuming a source-RIS distance \(z_0\), the corresponding spot area at the RIS is
\[
A_a \sim (\theta_{\mathrm{source}} z_0)^2 \sim \frac{\lambda_{s}^2 z_0^2}{A_{\mathrm{Tx}}}.
\]
Consequently, the fraction of source power captured by the RIS is
\[
\frac{L_{\mathrm{RIS}}^2}{A_a} \approx \frac{L_{\mathrm{RIS}}^2 A_{\mathrm{Tx}}}{\lambda_{s}^2 z_0^2},
\]
where \(L_{\mathrm{RIS}}\) represents the effective lateral aperture of the RIS.

\textbf{RIS-to-receiver channel:}
The main-lobe angular width of the reflected beam is approximated as
\begin{equation*}
\theta_{\mathrm{beam}} \sim \frac{\lambda_{s}}{L_{\mathrm{RIS}}}.
\end{equation*}
At a propagation distance between the RIS and the receiver, denoted by $z_1$, the main-lobe spot area becomes
\begin{equation*}
A_b \sim (\theta_{\mathrm{beam}} z_1)^2 \sim \left( \frac{\lambda_{s} z_1}{L_{\mathrm{RIS}}} \right)^2.
\end{equation*}

Thus, the total channel efficiency for mode $s$ is modeled as
\begin{equation*}
\xi_s = \frac{L^2_{\mathrm{RIS}}}{A_{a}}\cdot\frac{A_{\mathrm{Rx}}}{A_b} \approx \frac{L^2_{\mathrm{RIS}} A_{\mathrm{Tx}}}{\lambda_{s}^2 z_0^2}\cdot \frac{A_{\mathrm{Rx}} L_{\mathrm{RIS}}^2}{\lambda_{s}^2 z_1^2}=\frac{L^4_{\mathrm{RIS}}\cdot A_{\mathrm{Tx}}\cdot A_{\mathrm{Rx}}}{\lambda_{s}^4\cdot z_0^2 \cdot z_1^2},
\end{equation*}
where $A_{\mathrm{Rx}}$ denotes the effective aperture of the receiver.  
This captures the fundamental geometric loss due to beam spreading; additional RIS losses such as element efficiency or phase-quantization effects can be incorporated multiplicatively but are omitted here for simplicity.

A non-unit detection efficiency $\eta_s<1$ models losses arising from imperfect alignment at the receiver, such as spatial mode mismatch between the arrival signal and the detector or LO, polarization mismatch, and imperfect photon-to-electron conversion. These effects lead to signal attenuation and shot noises. Yet, excess additive noise sources such as dark counts and phase noise caused by laser linewidth and phase drift are neglected. Similar to $\xi_s$, $\eta_s$ is also wavelength-dependent. 

\subsection{Time-Resolving Quantum Measurement}
\begin{figure}[htpb]
	\centering
	\subfloat{
		\includegraphics[width=\linewidth]{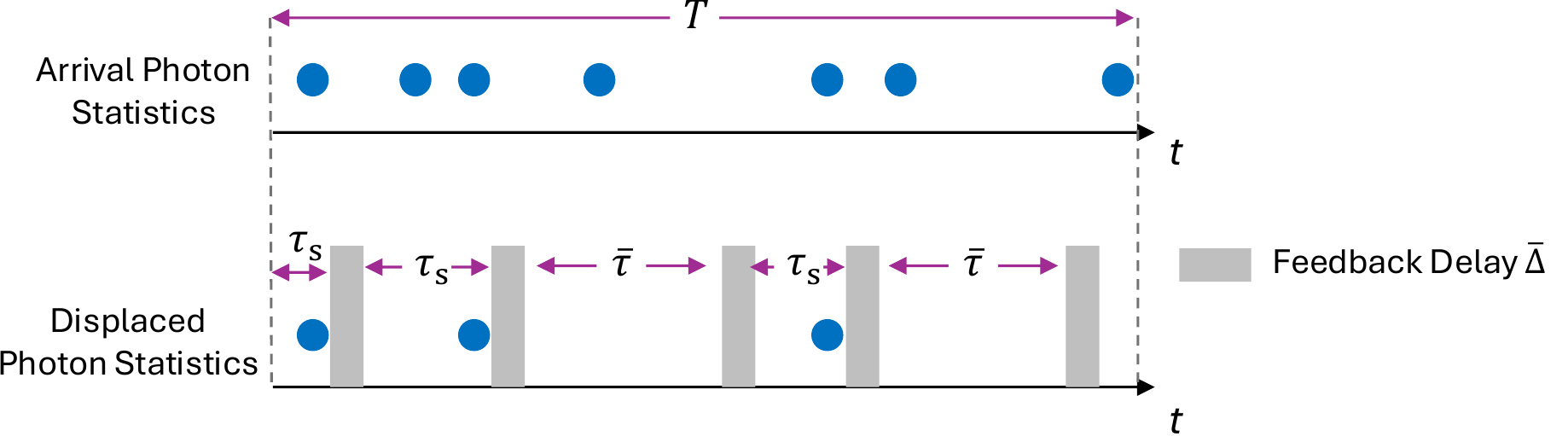}}
	\caption{An example showing an arrival light mode is adaptively displaced by the LO signal in 5 rounds, creating an increasingly sparse photon statistics in the SPD.}
    \label{fig:displaced:photon}
\end{figure}

The arrival multi-color light is first de-multiplexed and then independently handled and measured by the receiver. For the sake of simplicity, we elaborate the receiver design for one mode which is identical for all other modes. Specifically, given a LO signal $\beta$, the expected photon arrival rate can be modeled as a Poisson process with rate $\langle n(m,\beta) \rangle$, which is proportional to the overlap between the incoming mode $\alpha_m$ and LO $\beta$, and scaled by the area efficiency $\xi_s\eta_s$ according to Eq.(\ref{eq:mean-photon}). Photon arrival times are then simulated (or experimentally observed) as exponential waiting times with photon count $n$. 
Within each time bin of duration $\bar{\tau}$, the receiver collects the photon arrival times $\{\tau_s\}$ along with their corresponding mode indices $\{s\}$. Note that the receiver only marks the arrival time of the first detected photon within the time bin $\bar{\tau}$. An illustrative example is shown in Fig.\ref{fig:displaced:photon}. For one arrival mode of symbol duration $T$, the receiver makes 5 measurements, each of which applies a unique LO signal.

Next, the receiver performs Bayesian inference on photon detection events to adaptively select the optimal LO displacement that maximizes the posterior probability of the encoded coherent state. Algorithm~\ref{alg:adaptive} describes the proposed adaptive time-resolving quantum measurement. Initially, an unbiased estimator assigns an equal probability to all candidate symbols $\{\alpha_m\}_{m=1}^{M}$, i.e., $P_s(m) = 1/M$. After each measurement shot, the FPGA can update the hypothesis by calculating
\begin{equation}\label{eq:posterior}
P^{'}_s(m) =
\frac{P_s(m) \, \mathcal{\aleph}_s(m)}{\sum_{m'=1}^{M} P_{s}(m') \, \mathcal{\aleph}_{s}(m')},
\end{equation}
where $\mathcal{\aleph}_{s}(m)$ denotes the likelihood associated with the hypothesis that the input signal is $\alpha_m$, given the observed photon detection time. This likelihood accounts for both cases: (i) a single photon detected at time $\tau_s$, corresponding to $\mathcal{L}_s(\tau_s)$ in Eq.\eqref{eq:photon-arrival-time}, and (ii) no photon detected within a time bin of duration $\bar{\tau}$, as expressed below:
\begin{equation*}
\mathcal{\aleph}_{s}(m) =
\begin{cases}
e^{-\langle n(m,\beta) \rangle\, \bar{\tau}}, & \text{no click in} \,\, \bar{\tau}; \\[1ex]
\langle n(m,\beta) \rangle \, e^{-\langle n(m,\beta) \rangle\, \tau_s}, & \text{photon detected at time } \tau_s.
\end{cases}
\end{equation*}
If no photon is detected, the posterior is updated using the ``no-click'' likelihood model. Otherwise, the posterior is updated with each detected photon's timestamp and mode index. The FPGA calculates the updated hypothesis for all arrival modes according to Eq.(\ref{eq:posterior}). Then in the next measurement shot, the FPGA chooses the LO signal $\beta$ for the displacement operation corresponding to the current maximum a posteriori (MAP) hypothesis, i.e., 
\[
\beta = \alpha_{\hat{m}}, \quad 
\hat{m} = \hat{m}_{\hat{s}},
\]
where
\[
\hat{m}_s = \arg\max_m P_s(m), \quad 
\hat{s} = \arg\max_s P_s(\hat{m}_s).
\]

The posterior yielding the highest MAP probability among all modes is retained for the next iteration. 
The total observation time is incremented by the latest photon arrival plus a feedback delay $\bar{\Delta}$, and the adaptive measurement process continues until either the symbol duration $T$ is reached or the maximum number of measurement steps is performed. To ensure consistency between transmission and detection, the total measurement duration at the detector for each signal symbol is calibrated to match the symbol duration of the arrival modes. This alignment guarantees that the entire signal energy of each symbol is captured, prevents truncation or overlap of successive symbols, and preserves the photon statistics necessary for accurate adaptive measurements.

On a side note, to accelerate the LO adaptation when the displaced mean photon number is small, especially during the final few measurement steps (here, defined as those beyond $99\%$ of the symbol duration), we propose not to wait till the end of time bin $\bar{\tau}$ to make a Bayesian update. Rather, we will target for the first observed photon among all spectral modes and use its arrival time to calculate the Bayesian posteriror probability. 

\begin{algorithm}[t]
\caption{Adaptive Time-Resolving Quantum Measurement}
\label{alg:adaptive}
\begin{algorithmic}[1]
\Require Candidate states $\{\alpha_m\}_{m=1}^M$, true symbol $m^\star$, initial prior $P^{(0)}$, area efficiencies $\{\xi_s\eta_s\}$, symbol duration $T$, feedback delay $\bar{\Delta}$
\State $t \gets 0$, $P(m) \gets P^{(0)}$
\While{$t < T$}
    \State Select LO displacement $\beta = \alpha_{\hat{m}}$
    \State For each mode $s$, compute rate $n_{s}(m^\star, \beta)$
    \State Draw candidate photon arrival time for every mode 
    \State $\tau_s \sim {\rm Exponential}(n_s)$ 
    \State $\bar{\tau} \gets \min\{\bar{\tau}, T-t \}$
    \If{no photon arrives in $[0,\bar{\tau}]$ for mode $s$}
        \State $P_s(m) \gets \text{PosteriorUpdate}(P(m), \beta, \text{no-click}, \bar{\tau})$
        \State $t \gets t + \bar{\tau} + \bar{\Delta}$
    \Else
        \State Collect all photon timestamps $\{\tau_s\}$ with mode indices $\{s\}$ within the time bin $\bar{\tau}$
        \For{each detected photon $(\tau_s,s)$}
            \State $P_s(m)\gets \text{PosteriorUpdate}(P(m), \beta, \tau_s, s)$
        \EndFor
        \State $P(m) \gets \max_s P_s(m)$ 
        \State $t \gets t + \max_s \tau_s + \bar{\Delta}$
    \EndIf
\EndWhile
\State $\hat{m} \gets \hat{m}_{\hat{s}}$,\\ where $\hat{m}_s \!=\! \arg\max_m P_s(m)$, $\hat{s} \!= \!\arg\max_s P_s(\hat{m}_s)$
\State \Return $\hat{m}$ and final posterior $P(m)$
\end{algorithmic}
\end{algorithm}

\vspace{1ex}
\noindent\textbf{Computational Complexity:}
The proposed receiver scales linearly with the number of candidate states $M$, the number of spectral modes $S$, and the maximum number of measurement shots $T/\bar{\Delta}$. 
Each update shot requires computing $M$ photon rates and their corresponding likelihoods across $S$ modes, followed by a simple vector normalization. 
Therefore, the overall computational complexity is
\begin{equation*}
\mathcal{O}\big(M \,S \, T/\bar{\Delta}\big),
\end{equation*}
which remains tractable for real-time implementation with moderate $M$ and limited feedback latency. 
The algorithm thus provides a physically interpretable, computationally efficient, and hardware-compatible approach for adaptive time-resolving quantum measurement.

\section{Performance Evaluation}\label{sec:results}
In this section, we present simulation results to demonstrate the effectiveness of the proposed quantum-enhanced information retrieval system. The mean photon number of the probing coherent state for each mode is set as \( \langle n_0 \rangle = 1.5/S \), where \(S\) represents the total number of modes. The feedback delay $\bar{\Delta}$ is set to $1~\mu\mathrm{s}$.
\cite{burenkov2020time}. The system efficiency including the detection and channel efficiency for the central high-intensity light --- the mode falling within the infrared windows of the free space --- is $\xi\times\eta = 0.66$, while for other frequency components it is $\xi\times\eta = 0.46$. The maximum number of measurement steps is set as 200. The time bin of duration $\bar{\tau}$ is set as $T/10$, where $T$ is the symbol duration.
\begin{figure*}[htbp]
    \centering
    \subfloat[$M=2^4$, $R=2$]{
        \includegraphics[width=0.32\textwidth]{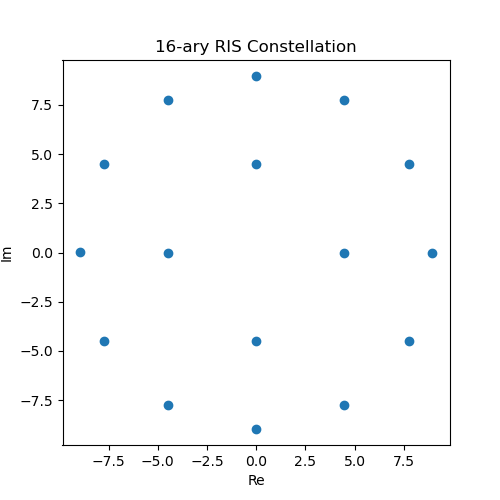}}
    \hfill
    \subfloat[$M=2^6$, $R=4$]{
        \includegraphics[width=0.32\textwidth]{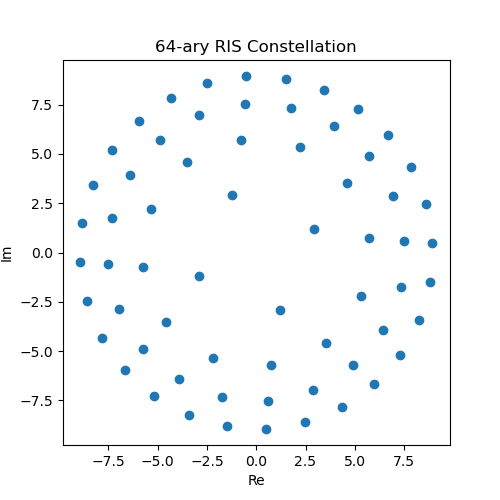}}
    \hfill
    \subfloat[$M=2^8$, $R=8$]{
        \includegraphics[width=0.32\textwidth]{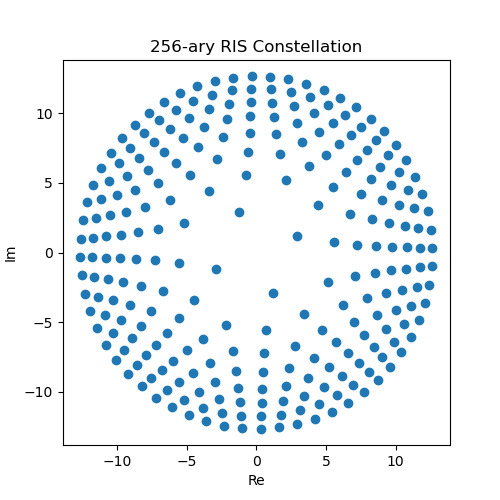}}
    \caption{Constellation of RIS modulation.}
    \label{fig:Constellation RIS modulation}
\end{figure*}

Simulation results are presented for $M$-ary modulations with $M = 2^4, 2^6, 2^8$, where the results are averaged over Monte Carlo trials of 20,000, 20,000, and 10,000, respectively. The corresponding interference visibility values $V$ considered for these modulations are 0.997, 0.998, and 0.9995, respectively. The constellations of the RIS-based $M$-ary modulations for $M = 2^4, 2^6, 2^8$ are illustrated in Fig.~\ref{fig:Constellation RIS modulation}.  

For a ring-based constellation design, $K_r$ denotes the intensity level of the $r$-th amplitude ring, and the number of phase states on the $r$-th ring is $l_r = 4(2r - 1)$,
ensuring that the number of phase samples increases linearly with the ring index, yielding finer phase resolution for outer rings. To maintain rotational symmetry and uniform angular spacing, an initial phase offset is introduced:
\begin{equation*}
\psi_{0,r} = \frac{\pi}{2 l_r}, \quad r = 1, 2, \dots, R,
\end{equation*}
where $R$ is the total number of amplitude rings. Each phase value on the $r$-th ring is then given by
\begin{equation*}
\psi_{k,r} = \psi_{0,r} + \frac{2 \pi (k-1)}{l_r}, \quad k = 1, 2, \dots, l_r.
\end{equation*}

For the amplitude rings:  
\begin{itemize}
    \item For $M = 2^4$, the amplitude rings are set as $K/4$ and $K$.  
    \item For $M = 2^6$, the amplitude of the $r$th ring $K_r$  is set as $(7r - 4)K/24$.  
    \item For $M = 2^8$, the amplitude of the $r$th ring $K_r$  is set as $(15r - 8)K/112$.  
\end{itemize}

\begin{figure*}[htbp]
    \centering
    \subfloat[$M=2^4$, $K=80$]{
        \includegraphics[width=0.32\textwidth]{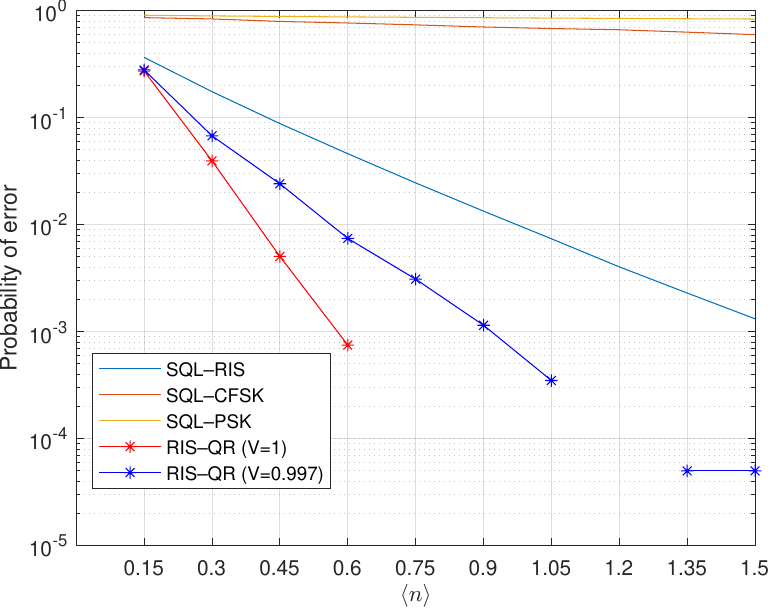}}
    \hfill
    \subfloat[$M=2^6$, $K=80$]{
        \includegraphics[width=0.32\textwidth]{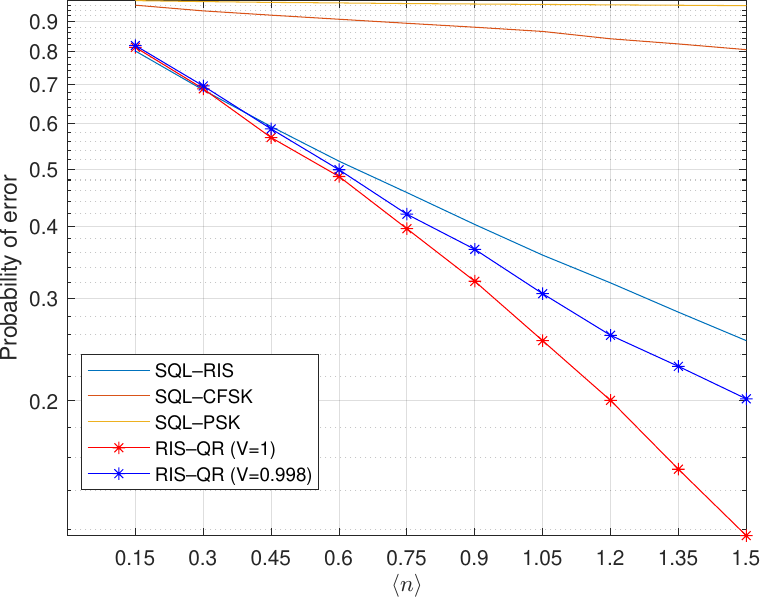}}
    \hfill
    \subfloat[$M=2^8$, $K=160$]{
        \includegraphics[width=0.32\textwidth]{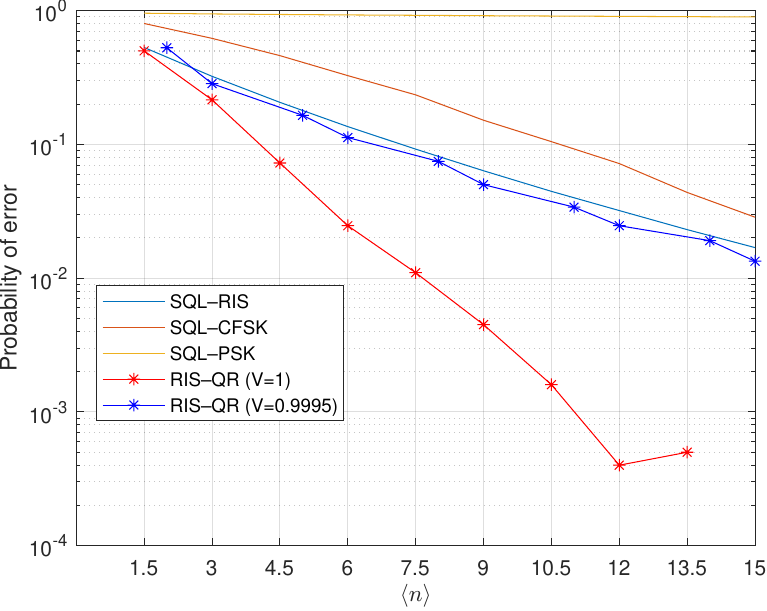}}
    \caption{Symbol error probability under different modulation schemes for $S=1$.}
    \label{fig:SQL varying n0}
\end{figure*}
Fig.~\ref{fig:SQL varying n0} shows the SQL for different modulation schemes. $\langle n \rangle$ represents the mean photon number of the input coherent state, i.e., the intensity of the probing light. We adopt two conventional modulation schemes --- phase shift keying (PSK) and combinatorial frequency shift keying (CFSK) --- for comparison \cite{burenkov2020time}. We observe that the classical detection methods for reading RIS-, PSK-, and CFSK-modulated symbols are limited by SQL for accuracy, among which RIS has the best performance. 
RIS-QR denotes the proposed quantum-enhanced system with an adaptive time-resolving quantum receiver. The blue (resp. red) curve with stars indicates its performance under imperfect (resp. perfect) visibility conditions. For fair comparison, we only use a single-mode probing light, i.e., $S=1$. The symbol pulse duration is set to $1,000~\mu$s.  It is observed that although imperfect visibility degrades the performance of the proposed quantum measurement protocol, the system still surpasses the SQL for $M = 2^4, 2^6, 2^8$-ary modulations across the board, demonstrating that quantum advantage is achieved regardless of the modulation order. Note that, due to the logarithmic scale on the y-axis, omitted dots in the curves represent zero. 

\begin{table*}[htbp]
\centering
\caption{Minimum $\langle n \rangle$ required to achieve the desired $P_e$}
\begin{tabular}{|c|c|c|c|c|c|}
\hline
$P_e$ &$V$ & $M$ & $\langle n \rangle$ by SQL-RIS & $\langle n \rangle$ by RIS-QR ($V<1$) & $\langle n \rangle$ by RIS-QR ($V=1$) \\
\hline
0.003 & 0.997  & $2^4$ & 1.35   & 0.9  & 0.6 \\
0.3   & 0.998  & $2^6$ & 1.35   & 1.2    & 1.05 \\
0.03  & 0.9995 & $2^8$ & 13.5   & 12   & 6 \\
\hline
\end{tabular}
 \label{table:desired Pe min n}
\end{table*}
Table~\ref{table:desired Pe min n} compares the minimum $\langle n \rangle$ required to achieve the desired $P_e$ using the optimal classical measurement (i.e., SQL-RIS) and the proposed quantum measurement under both imperfect and perfect visibility. It can be observed that the proposed quantum measurement always requires less energy (i.e., the minimum mean photon number $\langle n \rangle$)  to reach the desired $P_e$. Furthermore, under perfect visibility, the proposed quantum measurement requires only half the energy of the optimal classical measurement for $M = 2^4$ and $M = 2^8$. This also implies that, when using the same transmit energy and targeting the same accuracy, the communication distance can be extended by a factor of $\sqrt{2}$ compared with the classical measurement strategy, according to the channel model adopted in this work. In the following evaluations, we focus on the imperfect visibility ($V=0.9995$) to capture the realistic hardware constraints.

\begin{figure*}[htbp]
    \centering
    \subfloat[$M=2^4$, $T = 7 \mu s$]{
        \includegraphics[width=0.32\textwidth]{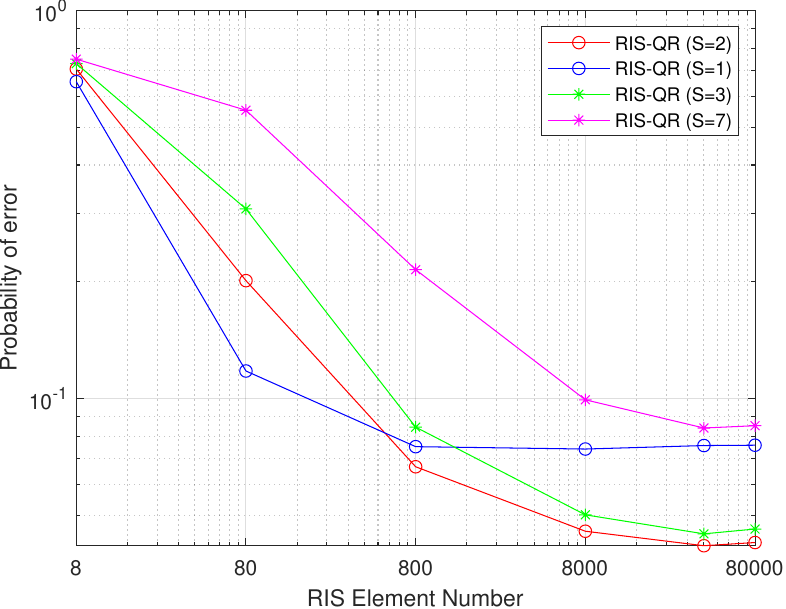}}
    \hfill
    \subfloat[$M=2^6$, $T = 15 \mu s$]{
        \includegraphics[width=0.32\textwidth]{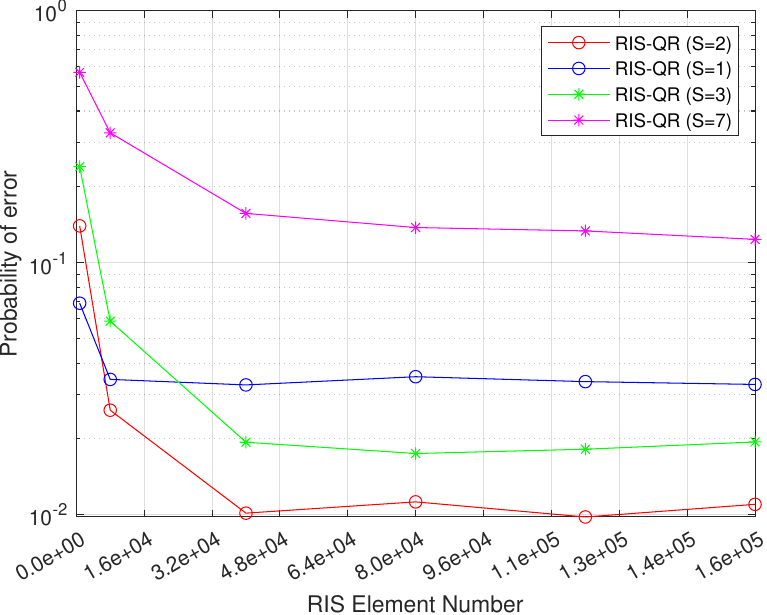}}
    \hfill
    \subfloat[$M=2^8$, $T = 30 \mu s$]{
        \includegraphics[width=0.32\textwidth]{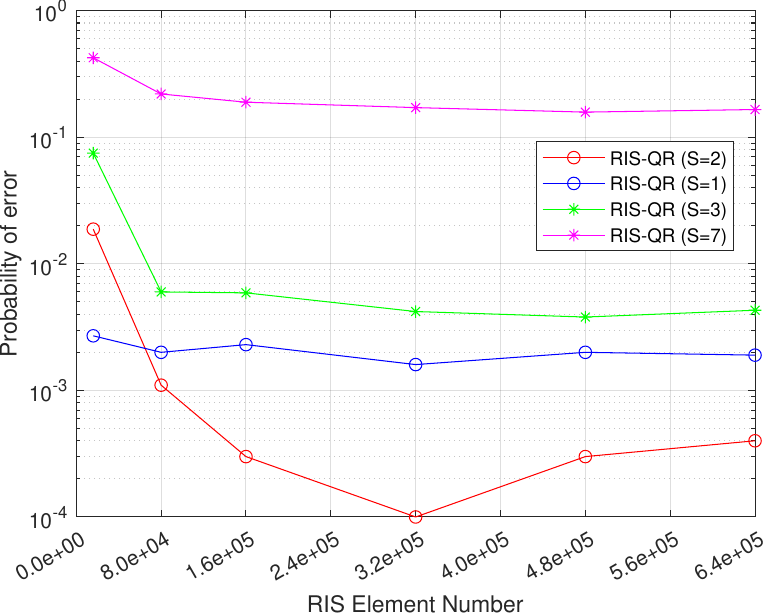}}
    \caption{Symbol error probability with varying RIS element number.}
    \label{fig:impact of RIS element number imperfect case}
\end{figure*}

\begin{figure*}[htbp]
    \centering
    \subfloat[$M=2^4$, $K=80*1000$]{
        \includegraphics[width=0.32\textwidth]{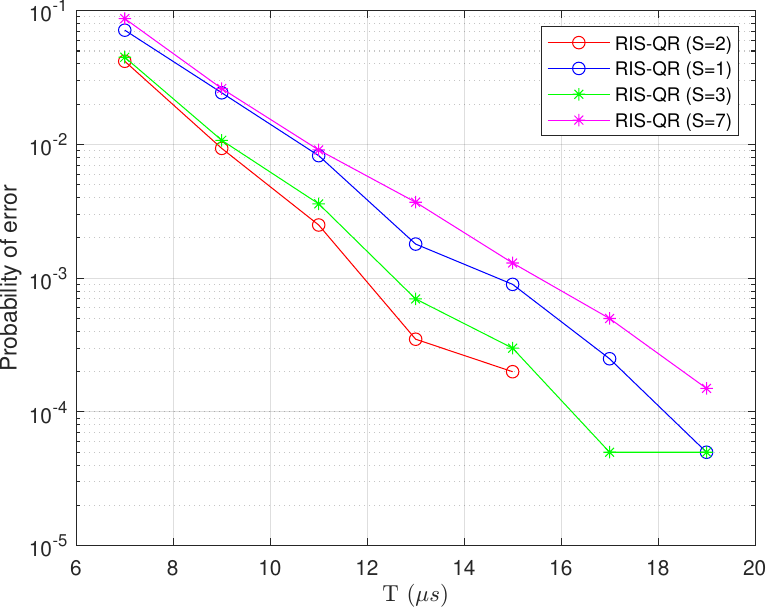}}
    \hfill
    \subfloat[$M=2^6$, $K=80*2000$]{
        \includegraphics[width=0.32\textwidth]{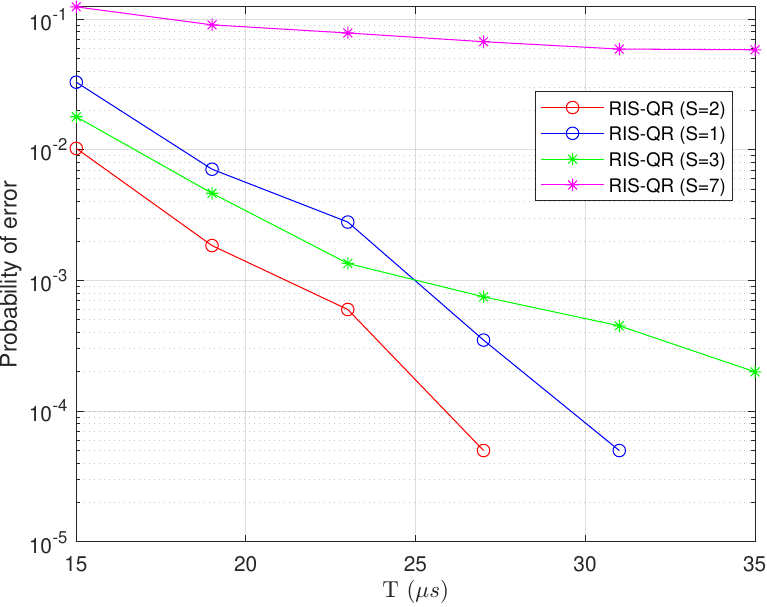}}
    \hfill
    \subfloat[$M=2^8$, $K=160*4000$]{
        \includegraphics[width=0.32\textwidth]{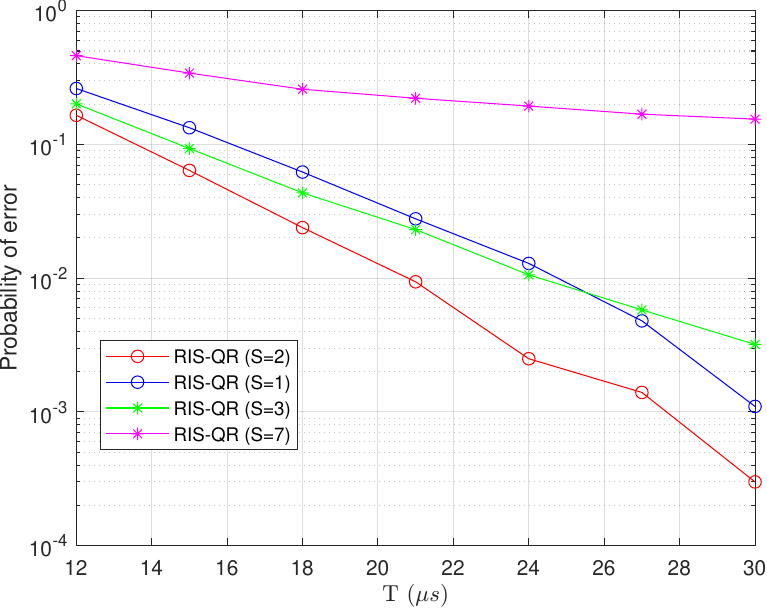}}
    \caption{Symbol error probability with varying symbol duration.}
    \label{fig:impact of symbol period imperfect}
\end{figure*}


Fig.~\ref{fig:impact of RIS element number imperfect case} compare the symbol error probability for different number of spectral modes $S$ as a function of the number of RIS elements. 
It can be observed that the system with a single probing mode, particularly when the RIS element number is small, achieves the best performance. This is because, at low photon intensities, dividing the energy among multiple modes significantly slows the photon arrival rate, thereby reducing the total information acquired within a limited measurement time. However, as the number of RIS elements increases, the photon intensity at the receiver enters the saturation regime~\cite{ingle2021passive,liu2022single}, due to RIS's beam focusing effect. In this regime, distributing the signal energy over more modes can enhance the acquired information by providing redundancy. Nevertheless, using too many modes (e.g., $S=7$) can degrade performance, particularly in challenging discrimination scenarios such as imperfect visibility or high-dimensional modulation tasks. This degradation arises for the same reason as at low photon intensities: the energy per mode becomes too small, limiting the photon arrival rate and thus reducing the overall information acquisition.

Fig.~\ref{fig:impact of symbol period imperfect} illustrates the symbol error probability $P_e$ as a function of the symbol period for $M = 2^4, 2^6, 2^8$-ary modulation. It is observed that, for the same transmit energy and RIS element number (i.e., the photon intensity at the receiver), longer symbol pulses result in superior discrimination performance. Moreover, even under imperfect visibility conditions, the proposed multi-color probing strategy with $S=2$ consistently achieves the lowest $P_e$, regardless of the measurement time budget (i.e., the symbol pulse duration).

\begin{figure*}[htbp]
    \centering
    \subfloat[$M=2^4$, $T = 13 \mu s$]{
        \includegraphics[width=0.32\textwidth]{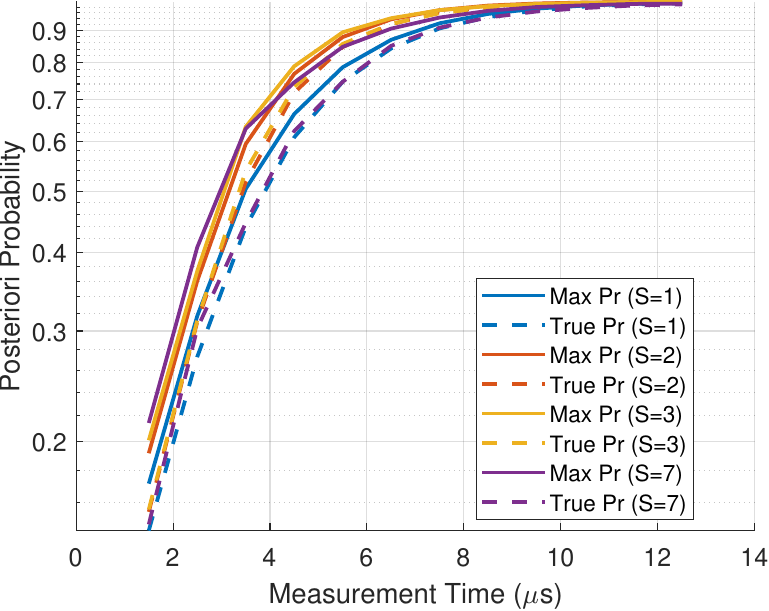}}
    \hfill
    \subfloat[$M=2^6$, $T = 23 \mu s$]{
        \includegraphics[width=0.32\textwidth]{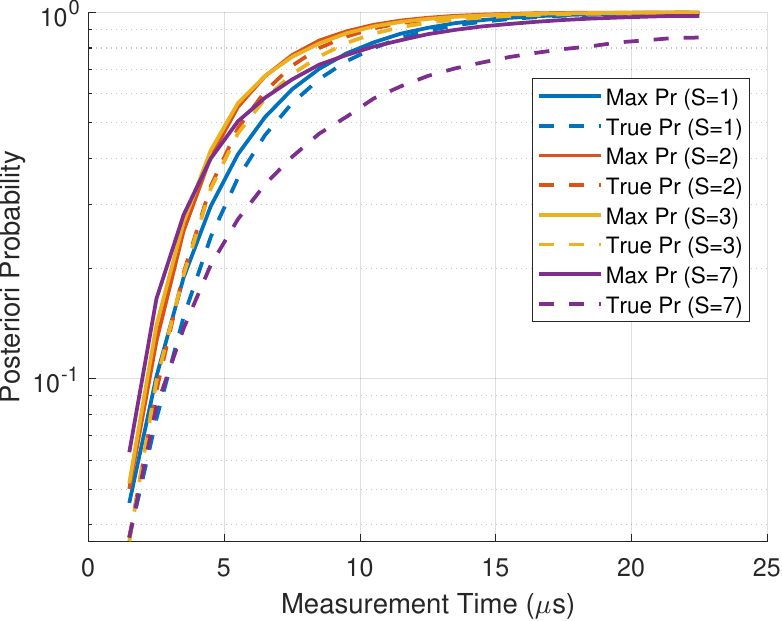}}
    \hfill
    \subfloat[$M=2^8$, $T = 30 \mu s$]{
        \includegraphics[width=0.32\textwidth]{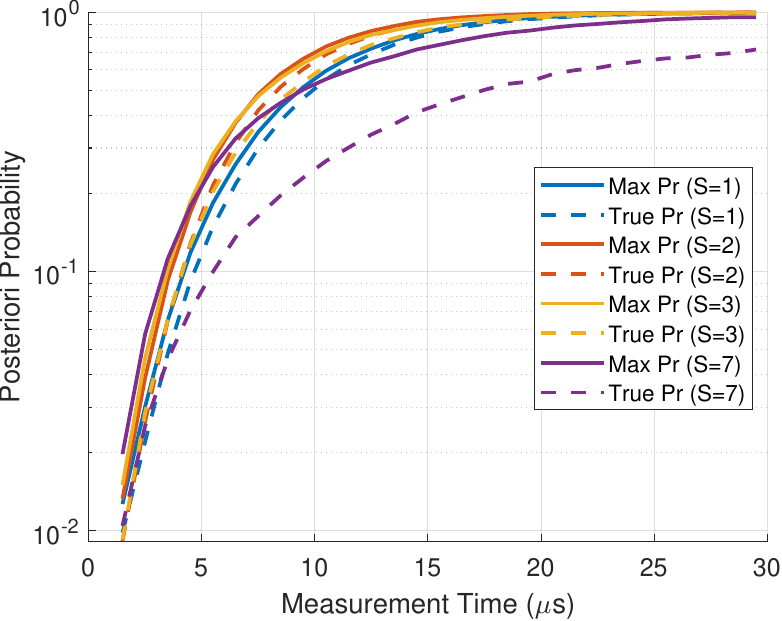}}
    \caption{Change of posteriori probability with varying measurement step.}
    \label{fig:Pe with measurement step imperfect}
\end{figure*}

Fig.~\ref{fig:Pe with measurement step imperfect} depicts the complete measurement procedure for a fixed symbol period. It implies the convergence rate of the Bayesian inference process. Here, ``Max Pr'' represents the maximum a posteriori probability over all possible symbols, while ``True Pr'' denotes the a posteriori probability of the true transmitted symbol. It is observed that using multiple spectral modes accelerates the increase in discrimination certainty. This confirms that dividing the transmit energy among multiple modes can enhance the information acquired during measurement, as a new adaptive LO is selected based on the MAP probability derived from photon arrival times across the multiple modes.

\begin{figure*}[htbp]
    \centering
    \subfloat[True Pr]{
        \includegraphics[width=0.45\textwidth]{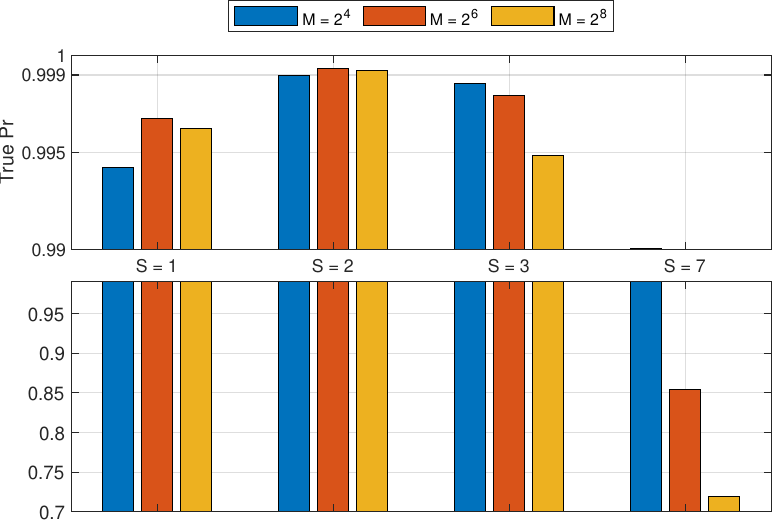}}
    \hfill
    \subfloat[Max Pr]{
        \includegraphics[width=0.45\textwidth]{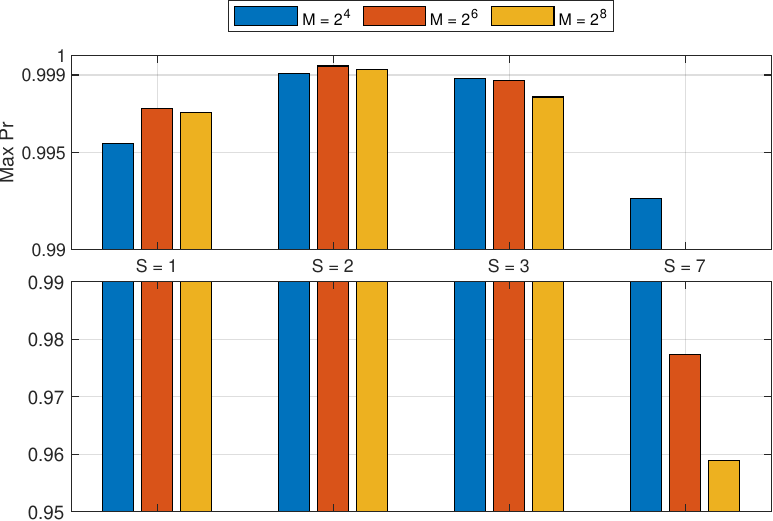}}
        
    \caption{Posteriori probability for different spectral modes at the final measurement step.}
    \label{fig:Pe at final step imperfect}
\end{figure*}

Fig.~\ref{fig:Pe at final step imperfect} shows the a posteriori probabilities corresponding to the final measurement time in Fig.~\ref{fig:Pe with measurement step imperfect}. It can be seen that, for multiple modes (i.e., $S=2$ and $S=3$), both the MAP probability and the a posteriori probability of the true transmitted symbol are consistently higher than those achieved with a single mode $S=1$. This observation is consistent with the phenomenon in Fig.~\ref{fig:impact of symbol period imperfect} that the symbol error probability $P_e$ for $S=2$ and $S=3$ is always lower than that for $S=1$ at a fixed symbol period.

\begin{figure*}[htbp]
    \centering
    \subfloat[$M=2^4$, $T = 13 \mu s$]{
        \includegraphics[width=0.32\textwidth]{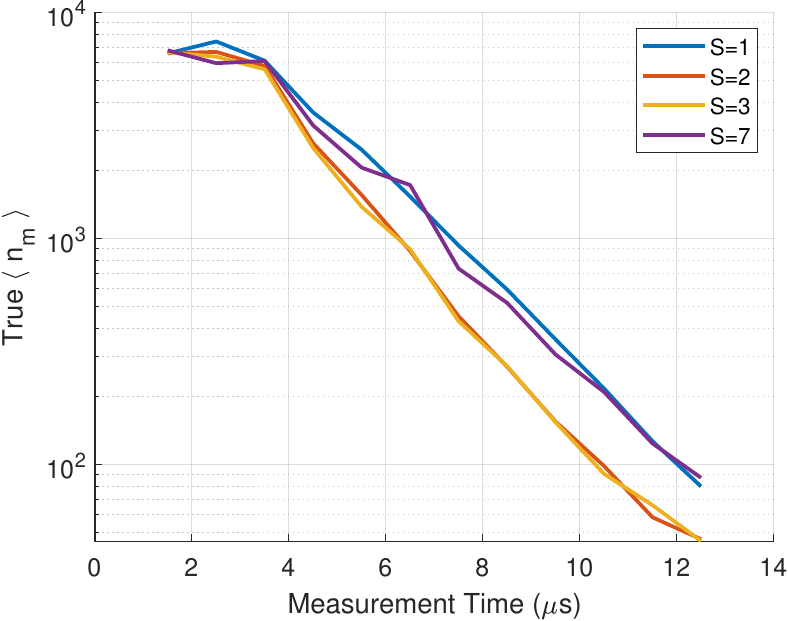}}
    \hfill
    \subfloat[$M=2^6$, $T = 23 \mu s$]{
        \includegraphics[width=0.32\textwidth]{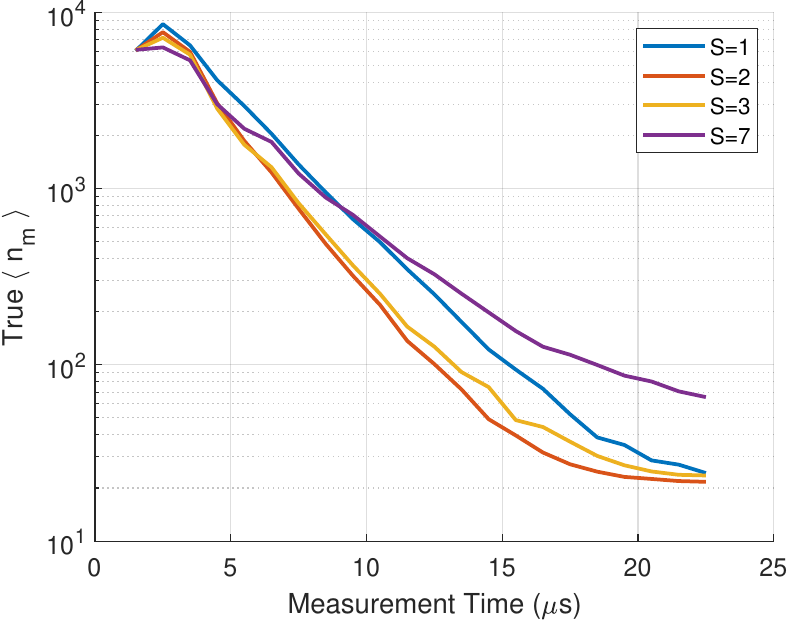}}
    \hfill
    \subfloat[$M=2^8$, $T = 30 \mu s$]{
        \includegraphics[width=0.32\textwidth]{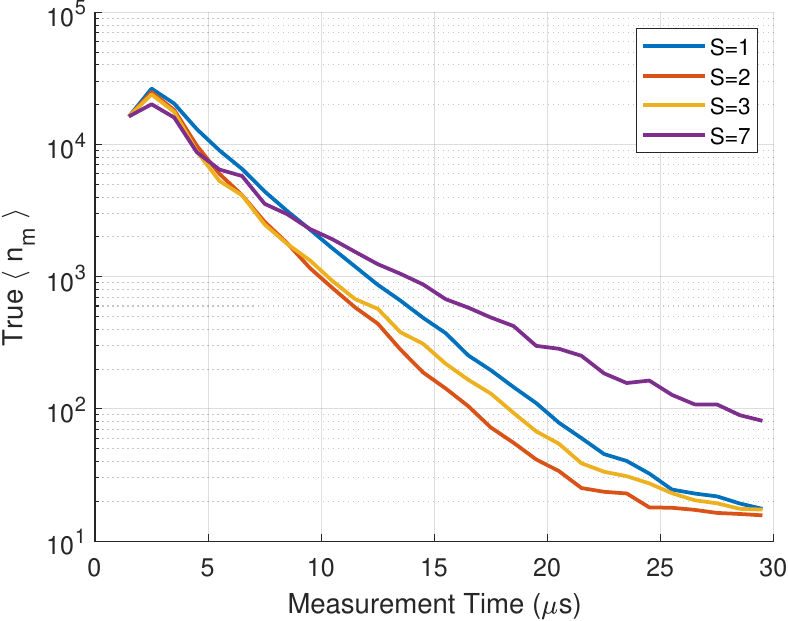}}
    \caption{Deviation between the estimated state ($\beta$) and the true state ($\alpha_m$) across varying measurement steps.}
    \label{fig:Deviation nnlambda with measurement step imperfect}
\end{figure*}
Fig.~\ref{fig:Deviation nnlambda with measurement step imperfect} shows the deviation between the estimated and true coherent states across measurement steps. 
To compare deviations across different modes while removing the effects of varying detection efficiency and input states, we define
\begin{equation*}
    \text{True }~\langle n_m \rangle
=
\frac{S}{T} \Big[
|\alpha_m|^2 + |\beta|^2 
- 2 V |\alpha_m| |\beta| \cos\big(\angle \alpha_m - \angle \beta \big)
\Big].
\end{equation*}
Using multiple modes can speed up and improve the estimation, particularly when $S=2$, independent of the modulation order $M$. For high-bit modulations (i.e., $M=2^6$ and $M=2^8$), further increasing the number of modes results in larger deviations at the end of the measurement steps, consistent with the observation that the $S=7$ case exhibits the worst estimation accuracy in Fig.~\ref{fig:impact of symbol period imperfect}.

\begin{figure*}[htbp]
    \centering

    \subfloat[$M=2^4$, $S=1$]{
        \includegraphics[width=0.23\textwidth]{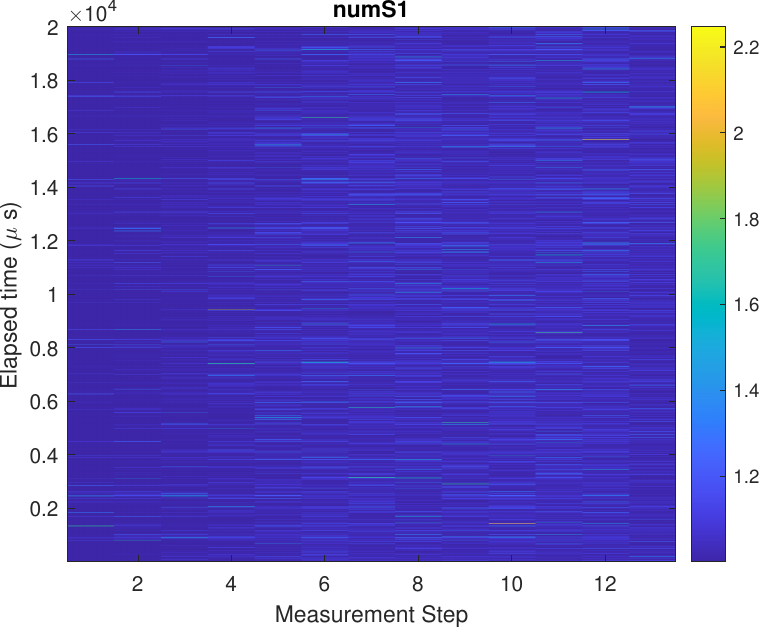}}
    \hfill
    \subfloat[$M=2^4$, $S=2$]{
        \includegraphics[width=0.23\textwidth]{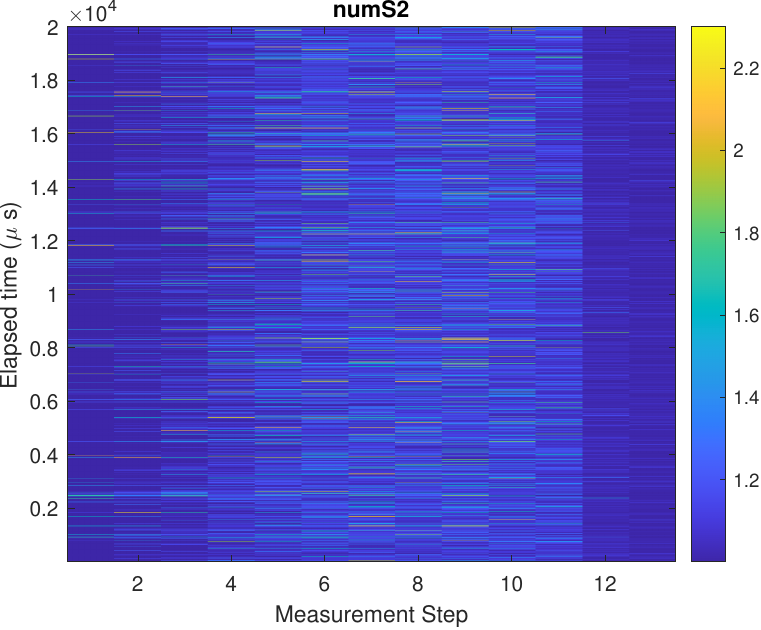}}
    \hfill
    \subfloat[$M=2^4$, $S=3$]{
        \includegraphics[width=0.23\textwidth]{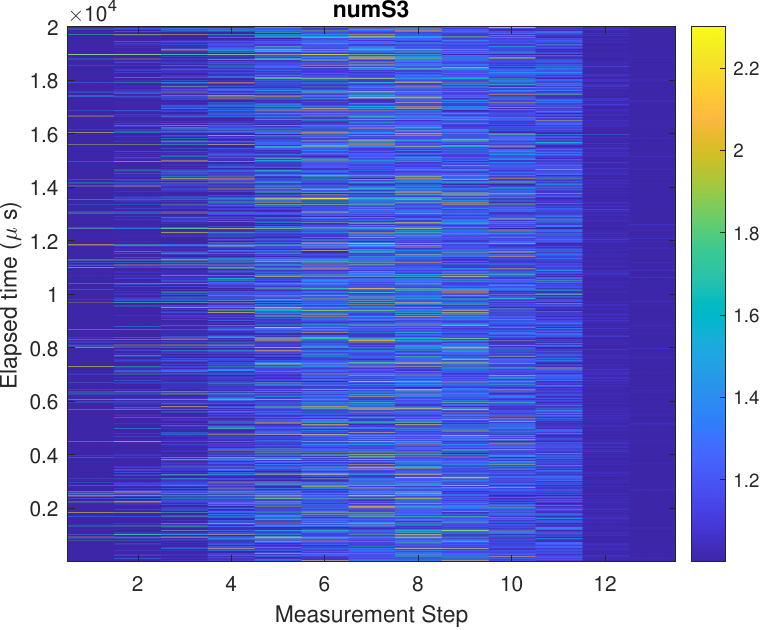}}
    \hfill
    \subfloat[$M=2^4$, $S=7$]{
        \includegraphics[width=0.23\textwidth]{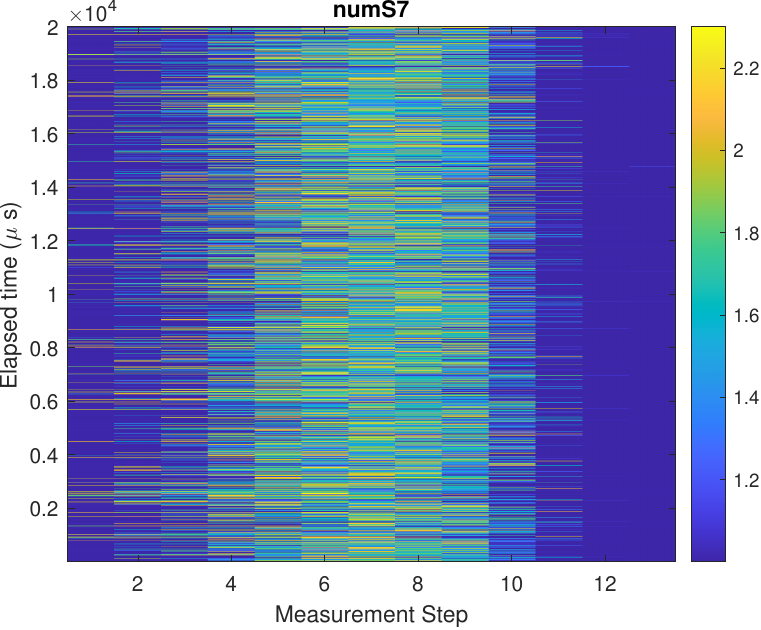}}

 \par\medskip   
    \subfloat[$M=2^6$, $S=1$]{
        \includegraphics[width=0.23\textwidth]{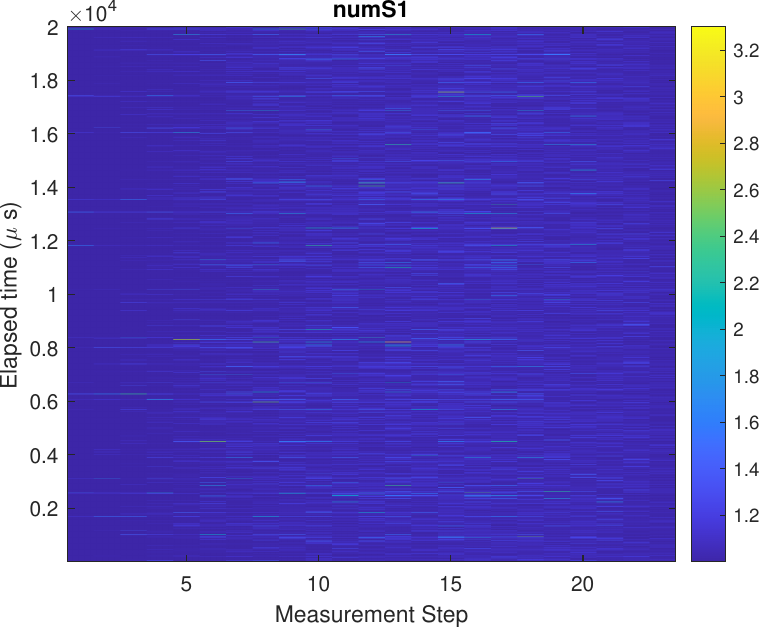}}
    \hfill
    \subfloat[$M=2^6$, $S=2$]{
        \includegraphics[width=0.23\textwidth]{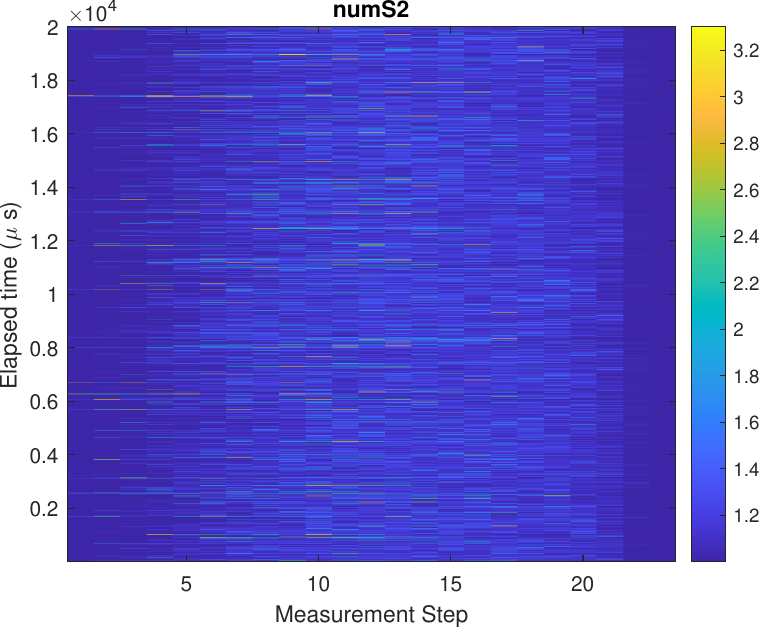}}
    \hfill
    \subfloat[$M=2^6$, $S=3$]{
        \includegraphics[width=0.23\textwidth]{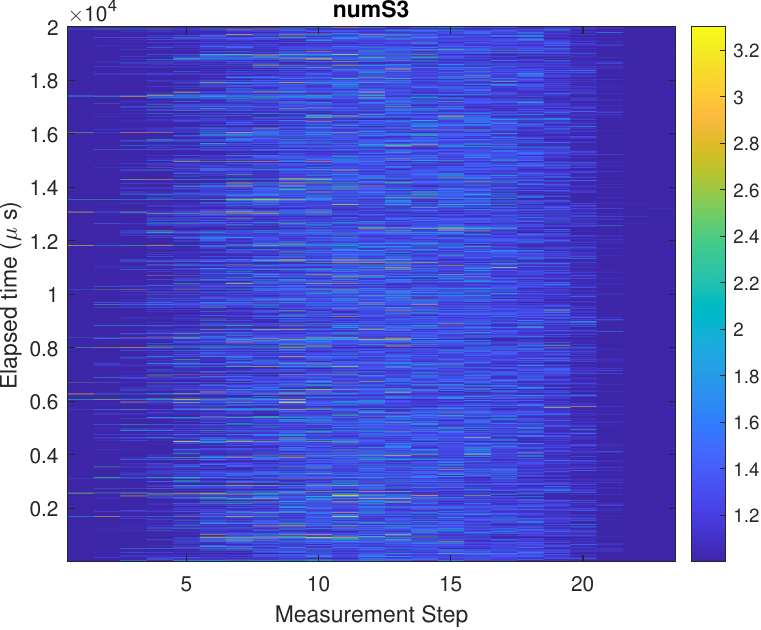}}
    \hfill
    \subfloat[$M=2^6$, $S=7$]{
        \includegraphics[width=0.23\textwidth]{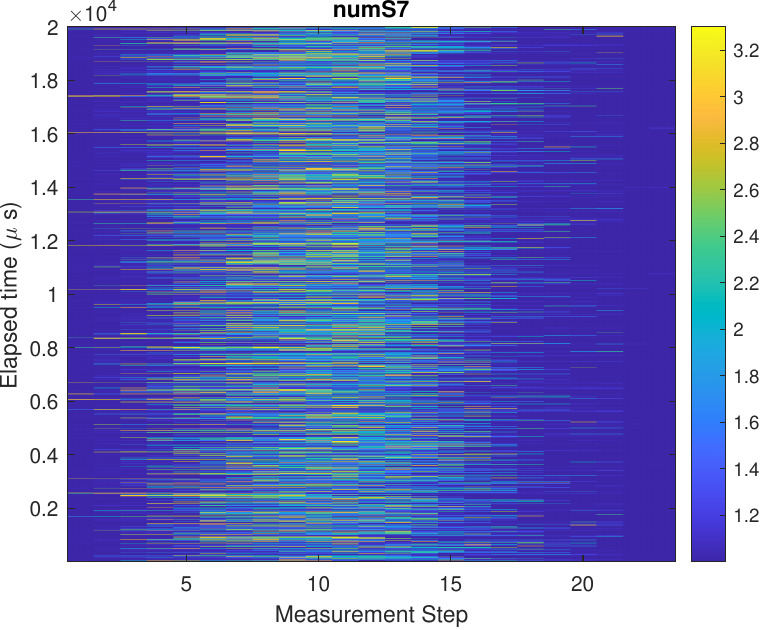}}
 \par\medskip   

    \subfloat[$M=2^8$, $S=1$]{
        \includegraphics[width=0.23\textwidth]{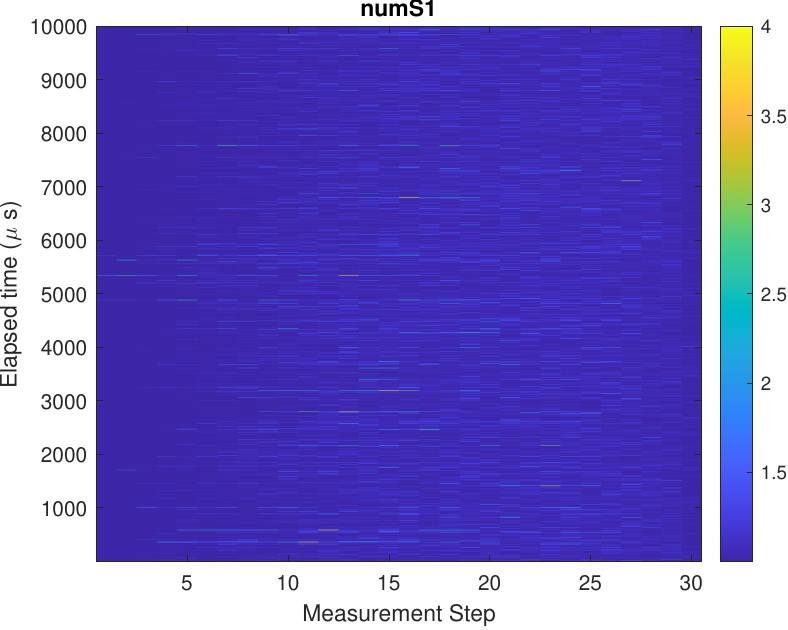}}
    \hfill
    \subfloat[$M=2^8$, $S=2$]{
        \includegraphics[width=0.23\textwidth]{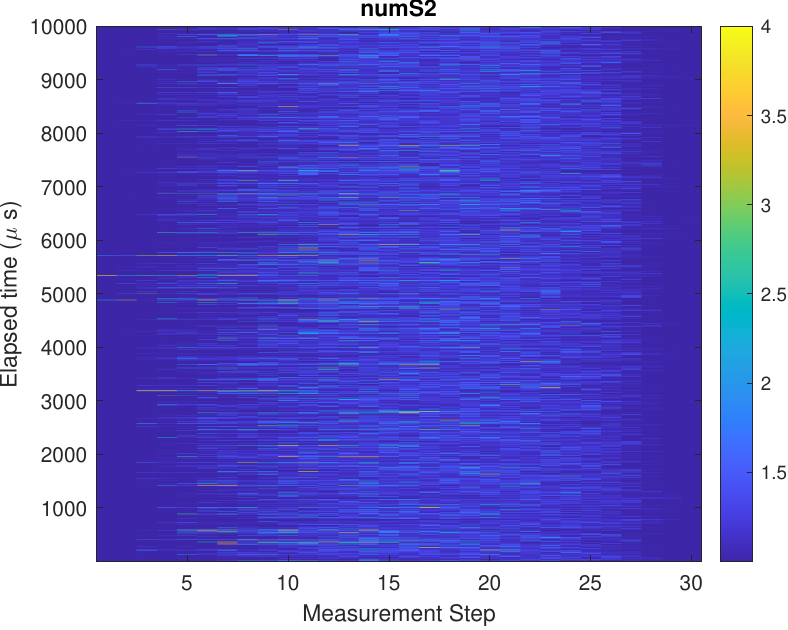}}
    \hfill
    \subfloat[$M=2^8$, $S=3$]{
        \includegraphics[width=0.23\textwidth]{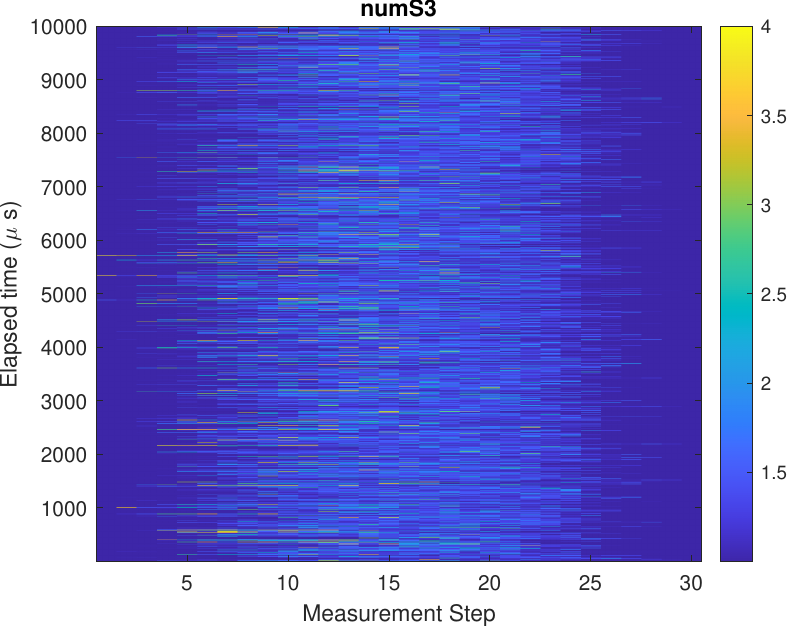}}
    \hfill
    \subfloat[$M=2^8$, $S=7$]{
        \includegraphics[width=0.23\textwidth]{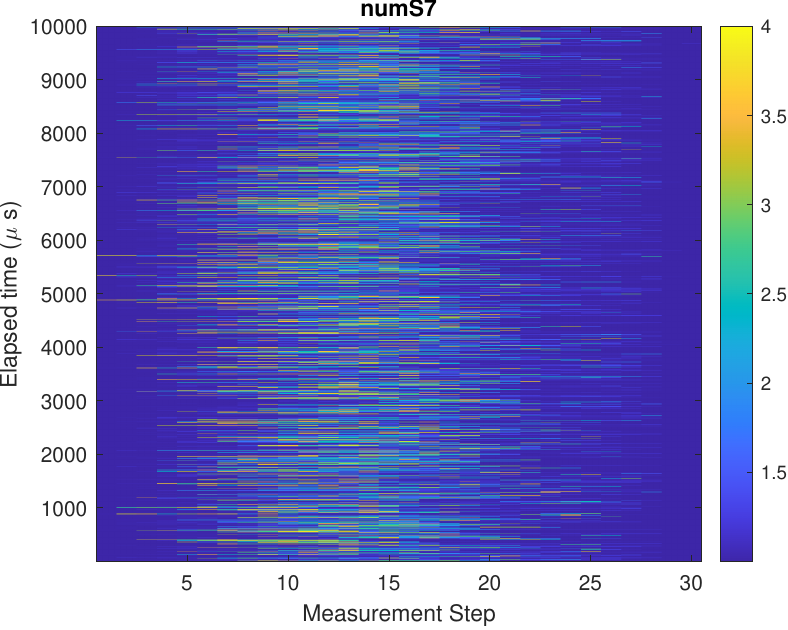}}

    \caption{Heatmap showing the elapsed time per measurement step for different modes $S \in \{1,2,3,7\}$. }
    \label{fig:Heatmap elapsed time imperfect}
\end{figure*}
Fig.~\ref{fig:Heatmap elapsed time imperfect} illustrates the distribution of elapsed time per measurement shot under imperfect visibility. It can be observed that when the number of modes increases to $S=7$, the occurrence of large elapsed times per measurement shot becomes more pronounced, which reduces the total information acquired within a limited measurement time. This effect arises because distributing the energy among multiple modes significantly slows the photon arrival rate, thereby limiting the total information that can be collected in the same time frame.

\begin{table*}[htbp]
\centering
\caption{Data rate for different M-ary modulations}
\begin{tabular}{|c|c|c|c|c|}
\hline
$V$ & $M$ & $T$ ($\mu$ s) & $S=1$, Data rate$/S$   (Mbps) & $S=2$, Data rate$/S$ (Mbps) \\
\hline
0.997 & $2^4$ & 15 & 0.2665    & 0.1333 \\
0.998 & $2^6$ & 23 & 0.2602    & 0.1303 \\
0.9995 & $2^8$ & 30 & 0.2664   & 0.1333  \\
\hline
1 & $2^4$ & 13 & 0.3076    & 0.1538 \\
1 & $2^6$ & 19 & 0.3158    & 0.1579 \\
1 & $2^8$ & 27 & 0.2963    & 0.1482 \\
\hline
\end{tabular}
 \label{table:date rate comparison}
\end{table*}
As shown in Table~\ref{table:date rate comparison}, the proposed quantum measurement scheme demonstrates strong robustness with respect to the modulation order, achieving approximately the same data rate for a fixed number of probing modes, i.e., $[(1-P_e)\log_2 M]/T$
for \(M = 2^4,\, 2^6,\) and \(2^8\)-ary modulations.
Moreover, considering the symbol error probabilities $P_e$ in Fig.~\ref{fig:impact of symbol period imperfect} for the best-performing case with $S=2$, we have $P_e = 2 \times 10^{-4}$ at $T=15~\mu$s for $M=2^4$, and $P_e = 3 \times 10^{-4}$ at $T=30~\mu$s for $M=2^8$. Under a common independent-error model, this implies that $M=2^8$-ary modulation achieves a lower block-error probability for any symbol block length. This confirms that the proposed scheme can attain superior block-error performance using high-order modulation (i.e., $M=2^8$). The metric ``Data rate$/S$'' denotes the average data rate per mode. As expected, when $S=2$ transmit modes are used, the average data rate per mode is approximately half of that achieved with $S=1$ transmit mode.

\section{Conclusion}\label{sec:conclude}
In this work, we presented a quantum-enhanced information retrieval technique for a RIS-based backscatter system. At the heart of this technique was an adaptive time-resolving quantum receiver along with a multi-color probing laser. By adaptively adjusting the receiver's LO signal, the photon statistics are manipulated to yield the best inference of the transmitted symbol. The evaluation results verified that the proposed technique achieved symbol error probabilities below the classical SQL for modulation sizes up to $M = 2^8$, under realistic channel conditions and imperfect hardware. Two spectral modes are found sufficient to achieve this goal while balancing design complexity and overhead. Furthermore, the technique results in superior data throughput, lower transmission power, and longer reading distance compared to classical backscatter systems. These results confirm the significant potential of the proposed technique for future communication and sensing systems.

\bibliography{newRefer}

\begin{IEEEbiography}[{\includegraphics[width=1in,height=1.25in,clip,keepaspectratio]{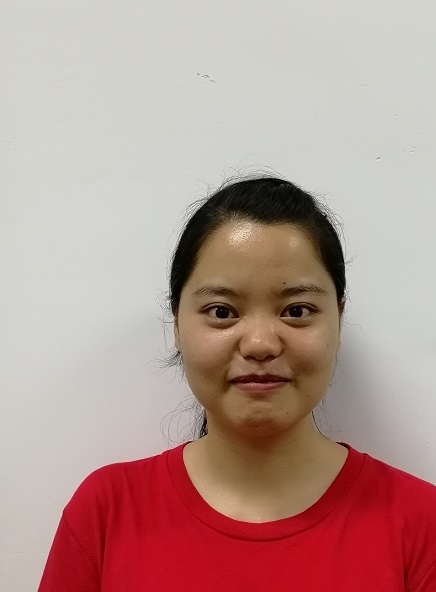}}]{Shiqian Guo}
     received the Ph.D. degree in information and communication engineering from South China University of Technology, in 2024. She is currently a Post-Doctoral Research Fellow with the
    Department of Computer Science, North Carolina State University. Her current research interests include wireless communications, federated learning, and quantum sensing.
\end{IEEEbiography}

\begin{IEEEbiography}[{\includegraphics[width=1in,height=1.25in,clip,keepaspectratio]{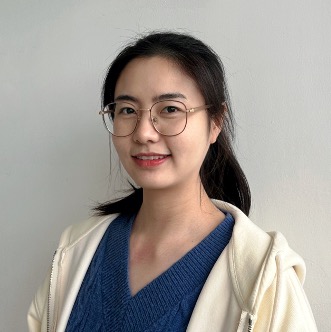}}]{Tingxiang Ji}
     received the B.S. and M.S. degrees in the Department of Computer Science and Technology at Nanjing Tech University. She is currently a PhD student with the Department of Computer Science, North Carolina State University. Her current research interests include quantum communications and graph theory.
\end{IEEEbiography}

\begin{IEEEbiography}
[{\includegraphics[width=1in,height=1.25in,clip,keepaspectratio]{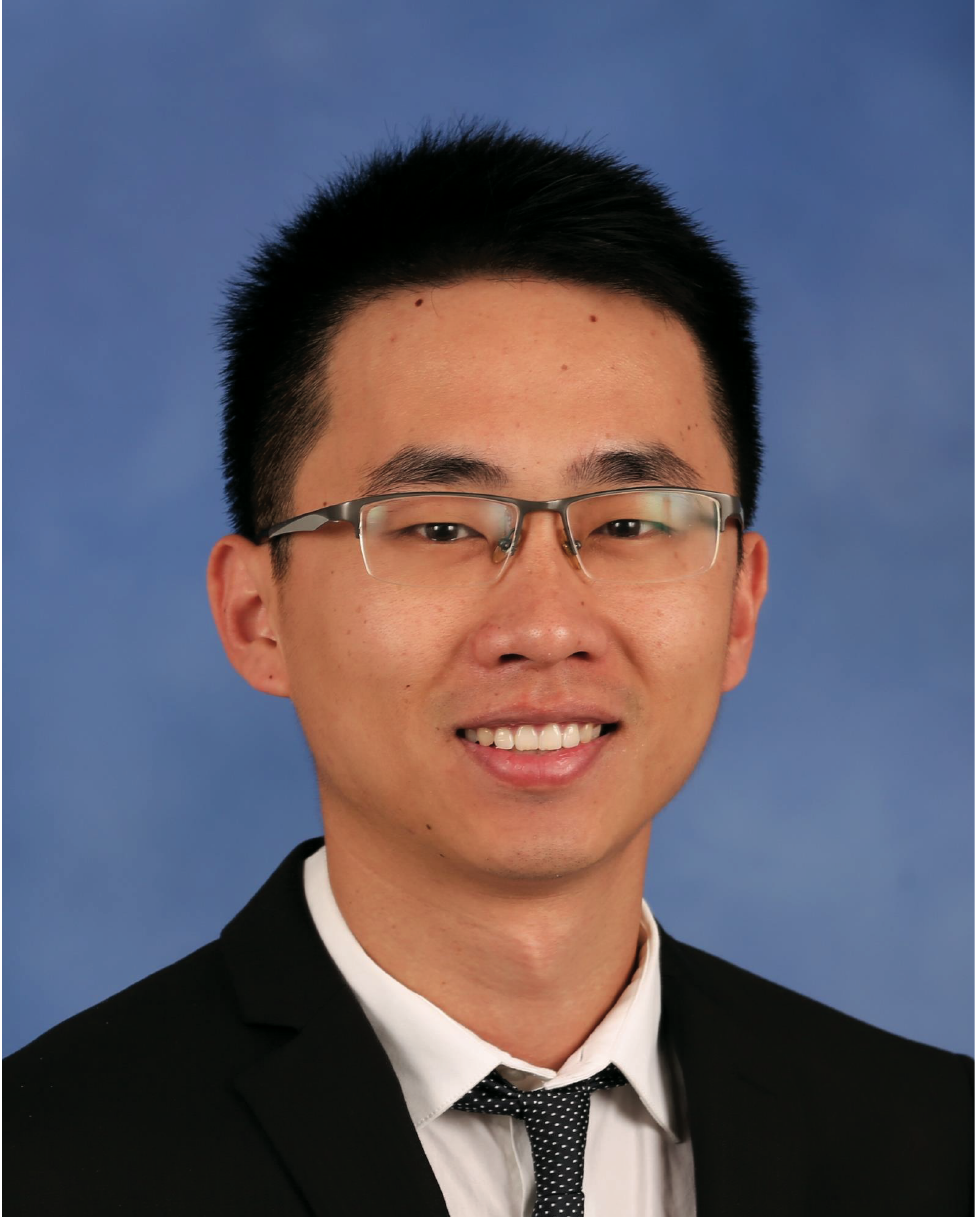}}]
	{Jianqing Liu}(Member, IEEE) is currently an associate professor with the Department of Computer Science at the NC State University. He received the Ph.D. degree from The University of Florida in 2018 and the B.S. degree from University of Electronic Science and Technology of China in 2013. His research interest is computer networks, quantum engineering, security and privacy. He received the US NSF CAREER Award in 2021 and several best paper awards including the 2018 Best Journal Paper Award from IEEE TCGCC. He currently serves as the associate editor with \emph{IEEE Transactions on Wireless Communications}, editor with \emph{IEEE Transactions on Communications}, and editor with \emph{IEEE Journal on Selected Areas in Communications - Quantum Series}. He was the associate editor with \emph{IEEE Transactions on Vehicular Technology} from 2021 to 2024.
\end{IEEEbiography}

\end{document}